\documentclass[lettersize,journal]{IEEEtran}
\usepackage{verbatim}
\ifCLASSINFOpdf
\else
   \usepackage[dvips]{graphicx}
\fi
\usepackage{multirow}
\usepackage{booktabs}
\usepackage{caption}
\usepackage{graphicx}
\usepackage{epstopdf}
\usepackage{subfigure}
\usepackage{amsmath}
\usepackage{amssymb}
\usepackage{amsthm}
\usepackage{amsfonts}
\usepackage[linesnumbered,boxed,ruled,commentsnumbered]{algorithm2e}
\usepackage{algpseudocode}
\usepackage{array}
\usepackage{cite}
\usepackage{url}
\usepackage{color}
\usepackage{epstopdf}
\usepackage{overpic}
\usepackage{lipsum}
\usepackage{cuted}
\usepackage{stfloats}
\usepackage{amsxtra}
\usepackage{threeparttable}
\usepackage{mathrsfs}
\usepackage{ulem}
\usepackage{makecell}
\usepackage{hyperref}
\usepackage[dvipsnames]{xcolor}
\hypersetup{
            colorlinks=true,
            linkcolor=black,
            anchorcolor=black,
            citecolor=green}
\begin{document}
\title{Interpretable Spectrum Transformation Attacks to Speaker Recognition}

\author{Jiadi Yao, Hong Luo, and Xiao-Lei Zhang
\thanks{Corresponding author: Xiao-Lei Zhang}
\thanks{Jiadi Yao, Xiao-Lei Zhang are with the School of Marine Science and Technology, Northwestern Polytechnical University, Xi'an 710072, China (e-mail: yaojiadi@mail.nwpu.edu.cn; xiaolei.zhang@nwpu.edu.cn).}
\thanks{Hong Luo is with the China Mobile (Hangzhou) Information Technology CO LTD, Hangzhou, China. (e-mail: luohong@cmhi.chinamobile.com).}
}

\markboth{Journal of \LaTeX\ Class Files,~Vol.~14, No.~8, August~2021}%
{Shell \MakeLowercase{\textit{et al.}}: A Sample Article Using IEEEtran.cls for IEEE Journals}


\maketitle

\begin{abstract}
The success of adversarial attacks to speaker recognition is mainly in white-box scenarios. When applying the adversarial voices that are generated by attacking white-box surrogate models to black-box victim models, i.e. \textit{transfer-based} black-box attacks, the transferability of the adversarial voices is not only far from satisfactory, but also lacks interpretable basis.
To address these issues, in this paper, we propose a general framework, named spectral transformation attack based on modified discrete cosine transform (STA-MDCT), to improve the transferability of the adversarial voices to a black-box victim model. Specifically, we first apply MDCT to the input voice. Then, we slightly modify the energy of different frequency bands for capturing the salient regions of the adversarial noise in the time-frequency domain that are critical to a successful attack.
Unlike existing approaches that operate voices in the time domain, the proposed framework operates voices in the time-frequency domain, which improves the interpretability, transferability, and imperceptibility of the attack. Moreover, it can be implemented with any gradient-based attackers. To utilize the advantage of model ensembling, we not only implement STA-MDCT with a single white-box surrogate model, but also with an ensemble of surrogate models. Finally, we visualize the saliency maps of adversarial voices by the class activation maps (CAM), which offers an interpretable basis to transfer-based attacks in speaker recognition for the first time. Extensive comparison results with five representative attackers show that the CAM visualization clearly explains the effectiveness of STA-MDCT, and the weaknesses of the comparison methods; the proposed method outperforms the comparison methods by a large margin.
\end{abstract}


\section{Introduction}
Speaker recognition is a task of identifying a person from voices that contain voice characteristics of the speaker \cite{reynolds1995robust,reynolds2000speaker,snyder2018x}, and finds its wide applications in real-world scenarios, such as bank trading, remote payment, and criminal investigations. State-of-the-art speaker recognition systems extract speaker embeddings with fixed-dimensions to represent the acoustic characteristics of speakers \cite{snyder2018x,bai2021speaker}. Prototypical speaker embeddings are i-vectors \cite{dehak2009support} extracted from Gaussian-mixture-model-based universal background models \cite{reynolds1995robust,reynolds2000speaker}. In recent years, with the fast development of deep learning, deep speaker embeddings become a new trend. Representative deep embeddings include d-vectors \cite{variani2014deep}, x-vector \cite{snyder2018x}, etc.


 Due to the wide applications of speaker recognition, its security is raising widespread concerns. Many attack techniques were developed to make speaker recognition systems easily failed, which can be categorized into three types: spoofing attacks, backdoor attacks, and adversarial attacks. There are mainly four sub-types of spoofing attacks: replay, voice conversion, impersonation and text-to-speech synthesis \cite{wu2015spoofing,das2020attacker}. Notably, the community-driven benchmark \textit{ASVspoof} Challenge series \cite{wu2017asvspoof} aim to address voice spoofing attacks. Recently, various detections \cite{cui2022synthetic,luo2021capsule} and spoofing countermeasures \cite{tak2021end} for spoofing attacks were developing rapidly. Backdoor attacks \cite{zhai2021backdoor,saha2020hidden,xu2022batt} provide poisoned data to the training data of a victim model in the training stage, and then activate the attack by presenting a specific small trigger pattern to the victim model in the test stage. Adversarial attacks aim to lead speaker recognition systems to wrong decisions by contaminating benign test examples with perceptually indistinguishable structured perturbations. The contaminated examples, a.k.a. adversarial examples, have been proven to significantly undermine deep-learning-based speaker recognition systems.

According to how much information a victim model can be accessed by an attacker, we categorize existing adversarial attacks into two settings, which are the white-box attacks and black-box attacks. In the \textit{white-box} setting, the attacker could access all information of the victim model, including the model architecture, parameters, and training data. For example, Villalba \textit{et al.} \cite{villalba2020x} demonstrated that gradient-based white-box attacks achieve high success rate to speaker verification systems. \textit{Black-box} attacks, which assume that the attackers know little about the victim models, are more practical and challenging than their white-box counterpart. Black-box attacks can be further partitioned to three sub-categories, which are the score-based, decision-based, and transfer-based attacks respectively. For the score-based attacks, the feedback from victim models is continuous, such as the posterior probability. For the decision-based attacks, the feedback is discrete, such as the recognition results. In both cases, a core problem is how to deal with the unknown gradients in the victim models. Existing solutions include gradient estimation and natural evolution. For example, Zhang \textit{et al.} \cite{9155483} proposed an adversarial example generation method named VMask, which estimates the gradients according to the difference between the similarity scores of multiple queries, and then uses zeroth-order optimization \cite{chen2017zoo} to solve the gradient-agnostic problem. Chen \textit{et al.} \cite{9519486} proposed the FakeBob attack to estimate the gradients through a natural evolutionary strategy \cite{ilyas2018black}. However, the above score-based and decision-based black-box attacks usually require a large number of queries.

To prevent large number of queries to victim models, \textit{transfer-based} black-box attacks were developed. Transfer-based black-box attacks \cite{villalba2020x,zhang2022nmi,9519486,zhang2020black} first generate adversarial examples from \textit{white-box surrogate models}, and then transfer the adversarial examples to \textit{black-box victim models}. Transfer-based methods do not utilize any information of the black-box victim models. Their success relies strongly on the transferability of the adversarial examples, which is the generalization ability of whether an adversarial example generated against a specific model can deceive other models. This paper focuses on discussing the transfer-based black-box attacks.


Although transfer-based black-box attacks for speaker recognition have received positive effects \cite{kreuk2018fooling,li2020adversarial,jati2021adversarial,villalba2020x}, the following core problems still need further investigation. (i) For a transfer-based attacker, transferability is a desired property of adversarial examples. However, it seems still far from explored. (ii) Many works generate adversarial examples in the time domain \cite{villalba2020x}, which may neglect the difference between the frequency bands of speech signals, while minor changes in frequency components may result in opposite decisions which is a well-known phenomenon. (iii) The success of an attacker cannot be interpreted straightforwardly, e.g. from the time-frequency spectrogram of a speech segment, which makes the design of an attack algorithm mysterious and heuristic.

To address the above issues, in this paper, we propose to improve the transferability of adversarial examples in the time-frequency domain by a novel framework, named \textit{spectrum transformation attack based on modified discrete cosine transform} (STA-MDCT). STA-MDCT {first performs MDCT on the input voice and then slightly modifies the energy of different frequency bands to alter the salient regions in the time-frequency domain that may lead the black-box victim model to an error decision.} To make the attack effect interpretable, we propose to visualize the saliency maps of adversarial examples via the class activation maps (CAM) \cite{li2022reliable}. By comparing the saliency maps of a voice before and after being added with an adversarial perturbation, we find that the adversarial examples generated with STA-MDCT are capable of shifting the attention of a black-box victim model, while its counterparts fail to do so, which provides an interpretable basis to transfer-based attacks.
To summarize, our main contributions are as follows:
 \begin{itemize}
 \item We propose the STA-MDCT framework. It is a general framework that any gradient based attacker can be applied with for probably improving its transferability to a black-box victim model. We implement two STA-MDCT variants, one with a single white-box surrogate model, and the other with an ensemble of white-box surrogate models.
 \item We propose to interpret the attack effect visually by CAM. To our knowledge, it is the first time that the transferability of an attacker can be interpreted visually beyond the final feedback from a victim model.
 \item We conducted adversarial attacks to four representative black-box victim speaker recognition systems. Extensive experiments demonstrate that the proposed STA-MDCT outperforms the state-of-the-art adversarial attack algorithms by a large margin in the transfer-based attack scenarios. The targeted attack success rate (TASR) can reach up to as high as 70.5\%.
 \end{itemize}

The rest of this paper is organized as follows. Section \ref{sec:related work} reviews the related work of adversarial attacks. Section \ref{sec:preliminaries} briefly describes some preliminaries. Section \ref{sec:methodology} describes the proposed method in detail. Section \ref{sec:Experimental Setup} shows the experimental setup, including the dataset, victim models, and evaluation metrics. Section \ref{sec:Result and Analysis} analyzes the experimental results. In Section \ref{sec:Conclusions}, we summarize the paper.

\section{Related Work}
\label{sec:related work}
In the following, we briefly make a literature survey of adversarial attacks in white-box and black-box scenarios.

White-box attacks can directly access the gradient of a victim model for generating adversarial examples. Since the first work named Fast Gradient Sign Method (FGSM) \cite{goodfellow2014explaining}, a large number of adversarial attack approaches have been proposed, including DeepFool \cite{moosavi2016deepfool}, I-FGSM \cite{kurakin2018adversarial}, PGD \cite{madry2017towards}, C\&W \cite{carlini2017towards}, and ACG \cite{yamamura2022diversified}. Besides, universal adversarial perturbations \cite{moosavi2017universal,xie2021real}, adversarial perturbations generative networks \cite{li2020universal, xie2021enabling, yao2023symmetric} are also extensively explored for generating real-time and efficient adversarial perturbations.

Black-box attacks can be categorized to score-based, decision-based, and transfer-based black-box attacks. For the score-based attacks, gradient-estimation \cite{chen2017zoo} and natural evolution strategies \cite{ilyas2018black,sharif2016accessorize} can be used to adapt perturbations to black-box victim models, given frequent queries to the victim models. For the decision-based attacks, boundary attack \cite{brendel2021decision} aims to find the best disturbance around invisible decision boundaries. \cite{chengquery} formulates the decision-based attack as a real-valued optimization problem which can be solved by any zeroth order optimization algorithm. HopSkipJumpAttack \cite{chen2020hopskipjumpattack} estimates the gradient directions at decision boundaries using Monte Carlo estimation.

Transfer-based attacks leverage the transferability of adversarial examples. Existing approaches that aim to improve the transferability can be categorized into four classes, which are (i) optimization-based algorithms, (ii) model augmentation strategy, (iii) ensemble learning \cite{zhang2022nmi} and meta-learning \cite{yuan2021meta} strategies, and (iv) modification of data distributions, respectively. The optimization-based algorithms aim to stabilize the optimization directions of adversarial perturbations, and avoid getting trapped in poor local optima of white-box surrogate models, e.g. Nesterov accelerated gradient \cite{zhang2022nmi,nesterov1983method}. The model augmentation strategy aims to simulate diverse models by applying loss-preserving transformations to inputs. For this purpose, Lin \textit{et al.} \cite{lin2019nesterov} utilize the scale-invariant property of deep neural networks to calculate the gradients over a set of images with different scales. Dong \textit{et al.} \cite{dong2019evading} optimize the perturbation over an ensemble of translated images to mitigate the issues of over-reliance on the surrogate model. The ensemble learning and meta-learning strategies generate transferable adversarial examples by integrating gradient information from multiple white-box surrogate models. Both of the strategies are beneficial in decreasing the gap between the surrogate models and the victim models. The approach of modifying data distributions aims to push the input data away from its original distribution for enhancing the adversarial transferability \cite{zhu2022toward}.

{The proposed STA-MDCT belongs to the second and fourth class of the above categories. Different from the above existing methods, we apply MDCT in the generation process of the adversarial voices for the first time, which improves the transferability and interpretability of adversarial voices in the time-frequency domain.}

\section{Preliminaries}
\label{sec:preliminaries}
\subsection{Speaker recognition}

This paper considers three representative subtasks of speaker recognition, including \textit{automatic speaker verification} (ASV), \textit{open-set identification} (OSI), and \textit{close-set identification} (CSI). We briefly present the definition of the subtasks as follows.

ASV aims to verify whether an anonymous utterance is pronounced by an enrolled person. A state-of-the-art speaker verification system first extracts a speaker embedding, e.g. x-vector, from an input utterance. Then, in the test phase, it calculates the similarity score between the speaker embeddings of an enrollment speaker $\mathbf{x}^{\mathrm{enroll}}$ and a test speaker ${\mathbf{x}}$. Finally, it compares the score with a predefined threshold $\theta$:
\begin{equation}
    {s}(\mathbf{x}^{\mathrm{enroll}},{\mathbf{x}}) \underset{H_{0}}{\stackrel{H_{1}}{\gtrless}} \theta,
\end{equation}
where ${s}(\mathbf{x}^{\mathrm{enroll}},{\mathbf{x}})$ is the similarity score between the two speakers, $H_{1}$ represents the hypothesis that ${\mathbf{x}}$ is uttered by the enrolled speaker, and $H_{0}$ is the opposite hypothesis of $H_{1}$.


Speaker identification aims to detect the speaker identity of a test utterance $\mathbf{x}$ from an enrollment database of $R$ speakers $\left\{r=\right.1,2, \cdots, R\}$ ($R>1$) by:
\begin{equation}
    r^{*}=\underset{r}{\operatorname{argmax}}\mbox{ }\{{s}(\mathbf{x}^{\mathrm{enroll}}_r,{\mathbf{x}})| \forall r = 1,\ldots,R\},
\end{equation}
where ${s}(\mathbf{x}^{\mathrm{enroll}}_r,{\mathbf{x}})$ denotes the similarity score between $\mathbf{x}$ and the $r$-th enrollment speaker $\mathbf{x}^{\mathrm{enroll}}_r$. If $\mathbf x$ can never be out of the $R$ enrolled speakers, then it is a CSI task; otherwise, it is an OSI task. From the above definitions, one can see that ASV is a special case of the OSI task with $R=1$. However, given the importance and wide applications of ASV in biometric authentication, this paper takes it as a separate task.

\subsection{White-box adversarial attack to speaker recognition}
\label{sec:baseline}
An adversarial attacker intentionally crafts a tiny perturbation $\delta$ that is indistinguishable by human, and then combines it with a benign voice $\mathbf{x}$ to produce a new one:
\begin{equation}
    \mathbf{x}^{\mathrm{adv}} = \mathbf{x} + \delta
\end{equation}
which may lead a victim model to wrong decisions according to the attacker's propose.
According to the goal of the attacker, adversarial attacks can be divided into \textit{targeted attacks} and \textit{untargeted attacks}.

Given a benign voice $\mathbf{x}$, the problem of searching for an adversarial example $\mathbf x^{\mathrm{adv}}$ can be formulated as solving the following constrained optimization problem:
\begin{align}\label{eq:general_loss}
    \max_{\mathbf{x}^{\mathrm{adv}}}&\mbox{ } \mathcal{L}\left(\mathbf{x}^{\mathrm{enroll}},\mathbf x^{\mathrm{adv}}\right)\\
    \mathrm{s.t.} & \mbox{ }  \left\|\mathbf x^{\mathrm{adv}}-\mathbf x\right\|_{p}<\epsilon
\end{align}
where $\mathcal{L}(\cdot)$ is a loss function that aims to make the victim model to errors, the $p$-norm $\left\|\mathbf x^{\mathrm{adv}}-\mathbf x\right\|_{p}$ represents the perturbation degree that controls the energy difference between the benign voice and the adversarial voice, which is upper-bounded by the predefined \textit{perturbation level} $\epsilon$. If $\epsilon$ was set small, then human may not be able to distinguish the adversarial voice $\mathbf x^{\mathrm{adv}}$ from the benign voice $\mathbf{x}$. In the following, we derive the definition of the adversarial attack for each speaker recognition subtask from  \eqref{eq:general_loss}.

To attack an ASV system, a targeted attack, a.k.a. \textit{impersonation attack}, aims to make the ASV system misclassify a non-target trial, i.e. $ {s}(\mathbf{x}^{\mathrm{enroll}},{\mathbf{x}}) <\theta$, to a target trial $ {s}(\mathbf{x}^{\mathrm{enroll}},{\mathbf{x}}^{\mathrm{adv}}) >\theta$, where a target trial indicates that the enrolled utterance and the test utterance belong to the same speaker identity, while a non-target trial is the opposite.
To achieve this goal, the loss function $\mathcal{L}_{\mathrm{imp}}$ can be defined as:
\begin{equation}
    \mathcal{L}_{\mathrm{imp}}\left(\mathbf{x}^{\mathrm{enroll}},\mathbf x^{\mathrm{adv}}\right)= {s}(\mathbf{x}^{\mathrm{enroll}},{\mathbf{x}}^{\mathrm{adv}})
\end{equation}

 In contrast, an untargeted attack, a.k.a. \textit{evasion attack}, aims to make the ASV system misclassify a target trial $ {s}(\mathbf{x}^{\mathrm{enroll}},{\mathbf{x}}) >\theta$  into a non-target trial $ {s}(\mathbf{x}^{\mathrm{enroll}},{\mathbf{x}}^{\mathrm{adv}}) <\theta$. Therefore, its loss function can be $\mathcal{L}_{\mathrm{eva}}$  defined as:
\begin{equation}
    \mathcal{L}_{\mathrm{eva}}\left(\mathbf{x}^{\mathrm{enroll}},\mathbf x^{\mathrm{adv}}\right)= -{s}(\mathbf{x}^{\mathrm{enroll}},{\mathbf{x}}^{\mathrm{adv}})
\end{equation}
 In this paper, we discuss the tasks of both the impersonation and evasion attacks to ASV.



 To attack a speaker identification system, a targeted attack aims to generate an adversarial voice such that the system may misclassify it as a target speaker; whereas an untargeted attack aims to lead the system to wrong predictions. Because a speaker identification system makes errors easily with the presence of natural noise interference which is similar to the effect from untargeted attacks, in this paper, we only discuss the tasks of the \textit{targeted attacks} to CSI and OSI. Suppose the attacker
aims to make the victim model wrongly predicts the $r$-th speaker into the $t$-th target speaker.
the objective of CSI is defined as:
\begin{align}
\label{equ:csi}
    \mathcal{L}_{\mathrm{CSI}}&(\mathbf{x}^{\mathrm{adv}},\{\mathbf{x}^{\mathrm{enroll}}_r\}_{r=1}^R)= {s}(\mathbf{x}^{\mathrm{enroll}}_t,\mathbf{x}^{\mathrm{adv}})\nonumber\\
   & -\max \{{s}(\mathbf{x}^{\mathrm{enroll}}_r,\mathbf{x}^{\mathrm{adv}})|\forall r = 1,\ldots,R, r\neq t\},
\end{align}

 and the objective of OSI is defined as:
\begin{align}
\label{equ:osi}
        \mathcal{L}_{\mathrm{OSI}}&(\mathbf{x}^{\mathrm{adv}},\{\mathbf{x}^{\mathrm{enroll}}_r\}_{r=1}^R)= {s}(\mathbf{x}^{\mathrm{enroll}}_t,\mathbf{x}^{\mathrm{adv}})\nonumber\\
   & -\max(\max \{{s}(\mathbf{x}^{\mathrm{enroll}}_r,\mathbf{x}^{\mathrm{adv}})|\forall r = 1,\ldots,R, r\neq t\},\theta).
\end{align}

where $\theta$ is a predefined threshold for the open-set problem of OSI.

In the following, we give a brief description of existing attackers, i.e. optimization algorithms, for the aforementioned adversarial attack objectives. For simplicity, we will omit $\mathbf{x}^{\mathrm{enroll}}$ from $\mathcal{L}(\cdot)$ in the rest of the paper, unless otherwise stated.
\subsubsection{Fast Gradient Sign Method (FGSM)}
FGSM \cite{goodfellow2014explaining} generates an adversarial example ${\mathbf x}^{\mathrm{adv}}$ by maximizing the loss $\mathcal{L}({{\mathbf x}^{\mathrm{adv}}})$ with one-step update to $\mathbf x$:
\begin{equation}
    \mathbf x^{\mathrm{adv}}=\mathbf x+\epsilon \cdot \operatorname{sign}(\nabla_{\mathbf x}\mathcal{L}(\mathbf x)),
\end{equation}
where $\nabla_{\mathbf x}\mathcal{L}(\mathbf x)$ is the derivative of the loss function with respect to $\mathbf x$.
\subsubsection{Iterative Fast Gradient Sign Method (I-FGSM)}
I-FGSM \cite{kurakin2016adversarial} extends FGSM to an iterative version with a small step size $\alpha$:
\begin{align}\label{eq:ifgsm}
    \mathbf{x}_{t+1}^{\mathrm{adv}}=\operatorname{Clip}_{{\mathbf x}, \epsilon}\left\{\mathbf{x}_{t}^{\mathrm{adv}}+\alpha \cdot \operatorname{sign}\left(\nabla_{\mathbf x_{t}^{\mathrm{adv}}} \mathcal{L}(\mathbf x_{t}^{\mathrm{adv}})\right)\right\}.
    \end{align}
where $\mathbf x_{0}^{\mathrm{adv}}=\mathbf x$, $t\in\{0,1,\cdots, T\}$ denotes the $t$-th iteration with $T$ as the maximum number of iterations, $\alpha = \epsilon / T$, and the function $\operatorname{Clip}_{{\mathbf x}, \epsilon}(\cdot)$ constrains the generated adversarial examples to be within the $\epsilon$-ball of $\mathbf x$ after each optimization step $t$.
\subsubsection{Momentum Iterative Fast Gradient Sign Method (MI-FGSM)}
MI-FGSM \cite{dong2018boosting} integrates the momentum into I-FGSM:
\begin{equation}
    \begin{array}{c}
    \mathbf{g}_{t+1}=\mu \cdot \mathbf{g}_{t}+\frac{\nabla_{\mathbf{x}_{t}^{\mathrm{adv}}} \mathcal{L}\left(\mathbf x_{t}^{\mathrm{adv}}\right)}{\left\|\nabla_{\mathbf{x}_{t}^{\mathrm{adv}}} \mathcal{L}\left(\mathbf{x}_{t}^{\mathrm{adv}}\right)\right\|_{1}}, \vspace{1ex} \\
    \mathbf{x}_{t+1}^{\mathrm{adv}}=\operatorname{Clip}_{{\mathbf x}, \epsilon}\left\{\mathbf{x}_{t}^{\mathrm{adv}}+\alpha \cdot \operatorname{sign}\left(\mathbf{g}_{t+1}\right)\right\}.
\end{array}
\end{equation}
where $\mathbf{g}_{0}=0$, $\mathbf{g}_{t}$ is the accumulated gradient at iteration $t$, and $\mu$ is the decay factor where $\mu = 1$ in our experiments.
\subsubsection{Nesterov Iterative Fast Gradient Sign Method (NI-FGSM)}
NI-FGSM \cite{lin2019nesterov} integrates Nesterov accelerated gradient into I-FGSM:
\begin{equation}
    \begin{array}{c}
    \mathbf{x}_{t}^{\mathrm{nes}}=\mathbf{x}_{t}^{\mathrm{adv}}+\alpha \cdot \mu \cdot \mathbf{g}_{t}, \vspace{1ex} \\
    \mathbf{g}_{t+1}=\mu \cdot \mathbf{g}_{t}+\frac{\nabla_{\mathbf{x}_{t}^{\mathrm{nes}}} \mathcal{L}\left(\mathbf x_{t}^{\mathrm{nes}}\right)}{\left\|\nabla_{\mathbf{x}_{t}^{\mathrm{nes}}} \mathcal{L}\left(\mathbf x_{t}^{\mathrm{adv}}\right)\right\|_{1}},\vspace{1ex}  \\
    \mathbf{x}_{t+1}^{\mathrm{adv}}=\operatorname{Clip}_{{\mathbf x}, \epsilon}\left\{\mathbf{x}_{t}^{\mathrm{adv}}+\alpha \cdot \operatorname{sign}\left(\mathbf{g}_{t+1}\right)\right\} .
\end{array}
\end{equation}
\subsubsection{Auto conjugate gradient attack}
Auto conjugate gradient attack (ACG) \cite{yamamura2022diversified} is based on conjugate gradient descent:
\begin{equation}
    \begin{aligned}
    \mathbf{y}_{t-1} & =\nabla_{\mathbf{x}_{t-1}^{\mathrm{adv}}} \mathcal{L}\left(\mathbf{x}_{t-1}^{\mathrm{adv}}\right)-\nabla_{{x}_{t}^{\mathrm{adv}}} \mathcal{L}\left(\mathbf{x}_{t}^{\mathrm{adv}}\right) \vspace{1ex} \\
    \beta^{HS}_{t} & =\frac{\left\langle-\nabla_{\mathbf{x}_{t}^{\mathrm{adv}}} \mathcal{L}\left(\mathbf{x}_{t}^{\mathrm{adv}}\right), \mathbf{y}_{t-1}\right\rangle}{\left\langle\mathbf{s}_{t-1}, \mathbf{y}_{t-1}\right\rangle} \vspace{1ex} \\
    \mathbf{s}_{t} & =\nabla_{\mathbf{x}_{t}^{\mathrm{adv}}} \mathcal{L}\left(\mathbf{x}_{t}^{\mathrm{adv}}\right)+\beta^{HS}_{t} \mathbf{s}_{t-1} \vspace{1ex} \\
    \mathbf{x}_{t+1}^{\mathrm{adv}} & =\operatorname{Clip}_{{\mathbf x}, \epsilon}\left\{\mathbf{x}_{t}^{\mathrm{adv}}+\eta_{t} \cdot \operatorname{sign}\left(\mathbf{s}_{t}\right)\right\}
\end{aligned}
\end{equation}
where the initial conjugate gradient $\mathbf{s}_{0}=0$, $\beta^{HS}$ is a parameter calculated from the past search information, the step size $\eta_t$ is dynamically adjusted and initialized by $\eta_0 = 2 \epsilon / T$. Particularly, when the number of iterations reaches a predefined value or the loss no longer drops, $\eta$ is halved.

From the above formulation, we see that ACG updates the {search points} in broader directions than the steepest {gradient} direction as that in FGSM, which may improve the transferability of the generated adversarial examples.

\section{Spectrum transformation attack based on modified discrete cosine transform}
\label{sec:methodology}
In this section, we first present the STA-MDCT framework in Section \ref{sec:framework}, and then describe the spectrum transformation in detail in Section \ref{sec:Spectrum Transformation}. Finally, in Section \ref{sec:Implementations}, we present two implementations of STA-MDCT, one with a single attacker and the other with an ensemble of attackers.

\begin{figure*}[t]
\begin{center}
\includegraphics[width=0.95\textwidth]{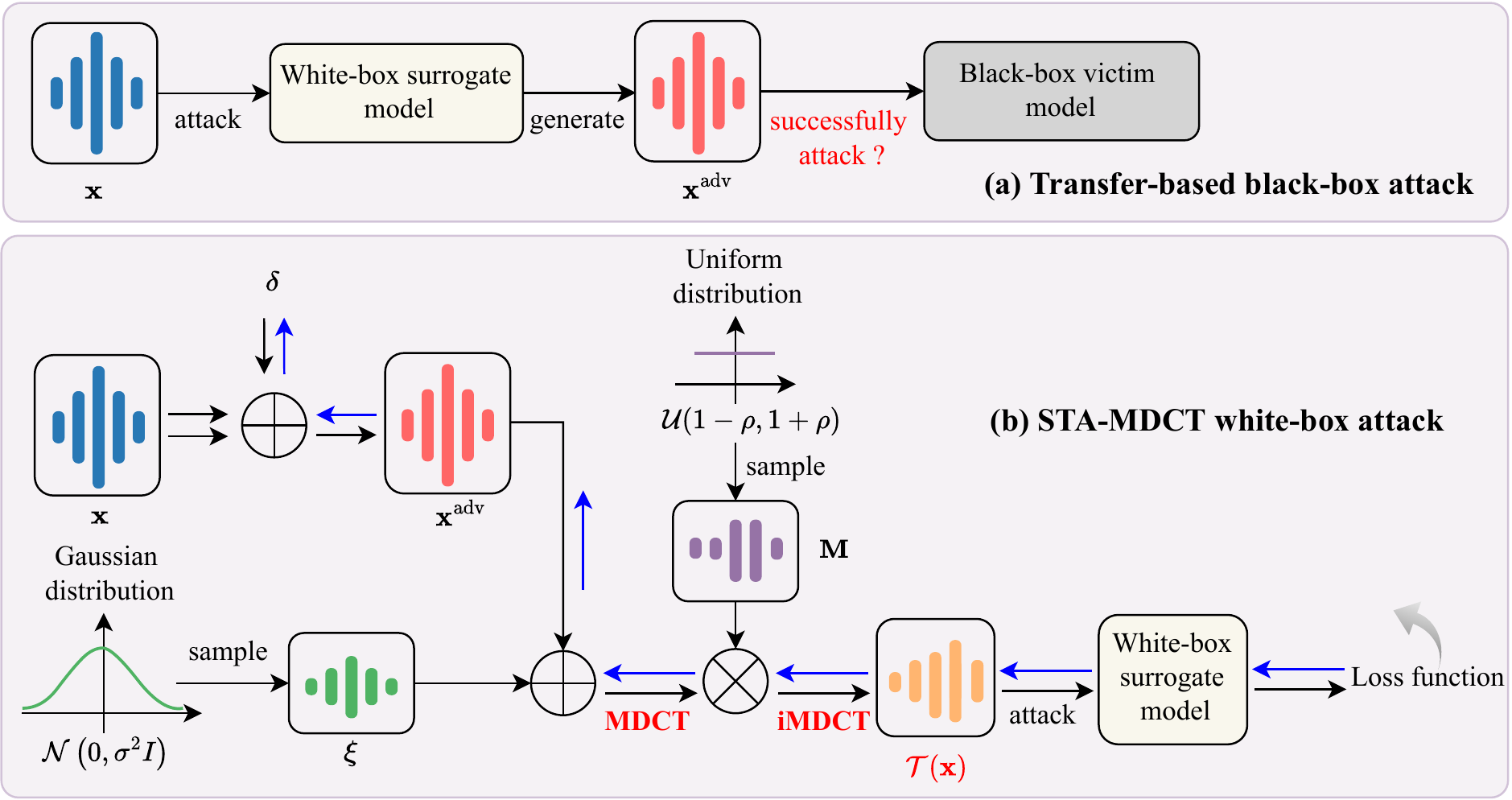}
\end{center}
\caption{{Framework of the STA-MDCT transfer-based attack, where $\mathbf{x}$ is a benign voice, $\mathbf{x}^{\mathrm{adv}}$ is an adversarial voice from $\mathbf{x}$, $\delta$ is the adversarial perturbation, $\xi$ is a random noise signal whose sample points are sampled from a Gaussian distribution $\mathcal{N}\left(0, \sigma^{2} I\right)$, $\mathbf{M}$ is a random matrix whose elements are sampled from a uniform distribution $ \sim\mathcal{U}(1-\rho, 1+\rho)$, and $\mathcal{T}(\mathbf x)$ is the spectrum transformation of $\mathbf x$.}}
\label{fig:STA-MDCT}
\end{figure*}



\subsection{Framework of the transfer-based STA-MDCT attack}
\label{sec:framework}
{Existing works usually apply loss-preserving transformations in the time domain, which might overlook the difference between frequency bands of speech signals, while slight variations in frequency components could lead to distinctly different decisions.} Given the same input, different victim models attend to different frequency bands and spectrum features of the input for making a decision \cite{wang2020high, li2022reliable}. Therefore, a perfect attacker trained from a white-box surrogate model in the time domain may have weak transferability to a black-box victim model that has different attention property in the time-frequency domain from the surrogate model.

To address this issue, the proposed STA-MDCT focuses on exploring the correlation between the victim models in the time-frequency domain.
Its framework is illustrated in Fig. \ref{fig:STA-MDCT}. As shown in Fig. \ref{fig:STA-MDCT}a, a transfer-based attacker first attacks a white-box surrogate model for generating transferable adversarial examples, and then uses them to attack a black-box target model. As shown in Fig. \ref{fig:STA-MDCT}b, for each optimization iteration, STA-MDCT first applies a spectrum transformation $\mathcal{T}(\mathbf x^{\mathrm{adv}})$ based on MDCT and inverse MDCT (iMDCT) to $\mathbf x^{\mathrm{adv}}$, and then reallocates the energy of the frequency bands of $\mathbf x^{\mathrm{adv}}$ according to the gradient information from the white-box surrogate model for improving the transferability of $\mathbf x^{\mathrm{adv}}$.

\subsection{Spectrum transformation}
\label{sec:Spectrum Transformation}

The spectrum transformation $\mathcal{T}(\mathbf x^{\mathrm{adv}})$ is defined as follows:
\begin{equation}
\label{equ:ST}
\begin{aligned}
    \mathcal{T}(\mathbf{x}^{\mathrm{adv}}) &=
    \mathrm{iMDCT}((\mathrm{MDCT}(\mathbf{x}^{\mathrm{adv}}) +\mathrm{MDCT}(\mathbf{\xi})) \odot \mathbf{M}), \\
    &=\mathrm{iMDCT}(\mathrm{MDCT}(\mathbf{x}^{\mathrm{adv}} +\mathbf{\xi}) \odot \mathbf{M})
\end{aligned}
\end{equation}
where $\xi$ is a random noise signal whose sample points are sampled from a Gaussian distribution $\mathcal{N}\left(0, \sigma^{2} I\right)$, $\mathbf{M}$ is a random matrix whose elements are sampled from a uniform distribution $ \sim\mathcal{U}(1-\rho, 1+\rho)$, the operator $\odot$ refers to the Hadamard product, the operator $\mathrm{MDCT}(\cdot)$ \cite{zhang2013mdct} is defined as:
\begin{equation}
    \begin{array}{c}
X_{\mathrm{MDCT}}(k)=\sum_{n=0}^{W-1} x(n) h(n)\cos \left[\frac{\left(2 n+1+\frac{W}{2}\right)(2 k+1) \pi}{2 W}\right], \\
\forall k=0,1, \ldots, \frac{W}{2}-1,\quad \forall n=0,1, \ldots, W-1
\end{array}
\end{equation}
 and the iMDCT operator $\mathrm{iMDCT}(\cdot)$ is defined as:
\begin{equation}
    \begin{array}{c}
x(n)=\frac{2}{W} h(n)\sum_{k=0}^{\frac{W}{2}-1} {X}_{\mathrm{MDCT}}(k) \cos \left[\frac{\left(2 n+1+\frac{W}{2}\right)(2 k+1) \pi}{2 W}\right], \\
\forall k=0,1, \ldots, \frac{W}{2}-1,\quad \forall n=0,1, \ldots, W-1
\end{array}
\end{equation}
with $h(n)$ denoted as the Kaiser-bessel-derived window.


Note that, MDCT is a linear orthogonal lapped transform, based on the time domain aliasing cancellation. In this paper, the adjacent frames produced by MDCT has an overlap of 50\%.

\subsection{STA-MDCT implementations}
\label{sec:Implementations}
Any gradient-based attackers can be applied to the proposed STA-MDCT framework. Different from \eqref{eq:general_loss}, the optimization objective of STA-MDCT is formulated as:
\begin{align}\label{eq:sta_loss}
    \max_{\mathbf{x}^{\mathrm{adv}}}&\mbox{ } \mathcal{L}\left(\mathbf{x}^{\mathrm{enroll}},\mathcal{T}(\mathbf{x}^{\mathrm{adv}})\right)\\
    \mathrm{s.t.} & \mbox{ }  \left\|\mathbf x^{\mathrm{adv}}-\mathbf x\right\|_{p}<\epsilon
\end{align}
Its optimization iteratively operates the following three steps:
\begin{itemize}
\item Calculate $\mathcal{T}(\mathbf x_{t}^{\mathrm{adv}})$ which is then back-propagated through the network to obtain the gradient information $\nabla_{\mathbf{x}_{t}^{\mathrm{adv}}} \mathcal{L}\left(\mathcal{T}(\mathbf x_{t}^{\mathrm{adv}}) \right)$, where $t$ denotes the $t$-th iteration.
\item Average $N$ gradients to obtain a more stable gradient direction.
\item Update the adversarial example $\mathbf{x}_{t+1}^{\mathrm{adv}}$ using an attacker algorithm, e.g. FGSM, I-FGSM, etc.
\end{itemize}

In this paper, we implement two STA-MDCTs, one with a single white-box surrogate model, and the other with an ensemble of white-box surrogate models.
\subsubsection{STA-MDCT based on a single surrogate model}
\label{sec:single-model STA-MDCT}
We apply I-FGSM \cite{kurakin2016adversarial} to STA-MDCT for attacking a white-box surrogate model. The optimization algorithm is formulated as:
\begin{equation}
\small
    \mathbf{x}_{t+1}^{\mathrm{adv}}=\operatorname{Clip}_{\mathbf{x}, \epsilon}\left\{\mathbf{x}_{t}^{\mathrm{adv}}+\\
    \alpha \cdot \operatorname{sign}\left(\frac{1}{N} \sum_{i=1}^{N} \nabla_{\mathbf{x}_{t}^{\mathrm{adv}}} \mathcal{L}\left(\mathcal{T}\left(\mathbf{x}_{t}^{\mathrm{adv}}\right)\right)\right)\right\}
\end{equation}
The algorithm is summarized in Algorithm \ref{algor1}.

\IncMargin{1em}
\begin{algorithm}[t]
\SetKwData{Left}{left}\SetKwData{This}{this}\SetKwData{Up}{up} \SetKwFunction{Union}{Union}\SetKwFunction{FindCompress}{FindCompress} \SetKwInOut{Input}{Input}\SetKwInOut{Output}{Output}\SetKwInOut{Return}{return}
	\Input{The enroll utterance ${\mathbf x}^{\mathrm{enroll}}$, the testing utterance ${\mathbf x}$ to be attacked, loss function $\mathcal{L}(\cdot)$, max iterations $T$, max perturbation $\epsilon$, step size $\alpha$, number of spectrum transformation $N$, tunning factor $\mathbf{\rho}$, std $\mathbf{\sigma}$ of noise $\mathbf{\xi}$.}
	\Output{An adversarial example ${\mathbf x}^{\mathrm{adv}}$.}
	 \BlankLine
	 $\mathbf x_{0}^{\mathrm{adv}}=\mathbf x$\;
	 \For{$t=0$ \KwTo $T-1$}{
	 	\For{$i=1$ \KwTo $N$}{\label{forins}
	 	Get spectrum transformation output $\mathcal{T}(\mathbf x_{t}^{\mathrm{adv}})$ using: $\mathcal{T}(\mathbf x)= \mathrm{iMDCT}(\mathrm{MDCT}(\mathbf x +\mathbf{\xi}) \odot \mathbf{M})$\;
	 	Gradient calculate: $\mathbf {k}_{i}= \nabla_{\mathbf{x}_{t}^{\mathrm{adv}}} \mathcal{L}\left(\mathcal{T}(\mathbf x_{t}^{\mathrm{adv}}) \right)$\;
	 	}
	 	Average gradient: $\mathbf {k}=\frac{1}{N} \sum_{i=1}^{N} \mathbf{k}_{i}$\;
	 	
	 	$\mathbf{x}_{t+1}^{\mathrm{adv}}=\operatorname{Clip}_{{\mathbf x}, \epsilon}(\mathbf{x}_{t}^{\mathrm{adv}}+\alpha \cdot \operatorname{sign}(\mathbf k))$\;
        }
	 ${\mathbf x}^{\mathrm{adv}}=\mathbf{x}_{T}^{\mathrm{adv}}$\;
	 \textbf{return} ${\mathbf x}^{\mathrm{adv}}$
    \caption{STA-MDCT}
    \label{algor1}
 	 \end{algorithm}
 \DecMargin{1em}

\subsubsection{STA-MDCT based on an ensemble of surrogate models (ensemble-STA-MDCT)}
\label{sec:ensemble}
To improve transferability, we apply I-FGSM to STA-MDCT for attacking an ensemble of white-box surrogate models (ensemble-STA-MDCT). The algorithm is summarized in Algorithm \ref{algor2}.

\IncMargin{1em}
\begin{algorithm}[t]
\SetKwData{Left}{left}\SetKwData{This}{this}\SetKwData{Up}{up} \SetKwFunction{Union}{Union}\SetKwFunction{FindCompress}{FindCompress} \SetKwInOut{Input}{Input}\SetKwInOut{Output}{Output}\SetKwInOut{Return}{return}
	\Input{White-box models $F=\left\{f_{1}, \ldots, f_{q}\right\}$, ensemble weight $w=\left[w_{1}, w_{2}, \ldots, w_{q}\right]$,
	the enroll utterance ${\mathbf x}^{\mathrm{enroll}}$, the testing utterance ${\mathbf x}$ to be attacked, loss function $\mathcal{L}_{f_{j}}(\cdot)$ for model $f_{j}$, max iterations $T$, max perturbation $\epsilon$, step size $\alpha$, number of spectrum transformation $N$.}
	\Output{An adversarial example ${\mathbf x}^{\mathrm{adv}}$.}
	 \BlankLine
	 $\mathbf x_{0}^{\mathrm{adv}}=\mathbf x$\;
	 \For{$t=0$ \KwTo $T-1$}{
	  \For{all $f_{j}$}{
	 	Compute the average gradient of the $N$ augmented models:
	 	$\mathbf{k}_{f_j}=\frac{1}{N} \sum_{i=1}^{N} \nabla_{\mathbf{x}_{t}^{\mathrm{adv}}} \mathcal{L}_{f_{j}}\left(\mathcal{T}\left(\mathbf{x}_{t}^{\mathrm{adv}}\right)\right)$
	 	}
	 	Fuse these gradients: $\mathbf {k}=\sum_{j=1}^{q} w_{j} \mathbf{k}_{f_j}$\;
	 	
	 	Update $\mathbf{x}_{t+1}^{\mathrm{adv}}$ by applying the sign gradient as:
	 	$\mathbf{x}_{t+1}^{\mathrm{adv}}=\operatorname{Clip}_{{\mathbf x}, \epsilon}(\mathbf{x}_{t}^{\mathrm{adv}}+\alpha \cdot \operatorname{sign}(\mathbf k))$\;
        }
	 ${\mathbf x}^{\mathrm{adv}}=\mathbf{x}_{T}^{\mathrm{adv}}$\;
	 \textbf{return} ${\mathbf x}^{\mathrm{adv}}$
    \caption{Ensemble-STA-MDCT}
    \label{algor2}
 	 \end{algorithm}
 \DecMargin{1em}

\section{Interpretability of STA-MDCT with saliency maps}
\label{sec:Interpretability}

In this section, we first introduce saliency maps in Section \ref{subsec:saliency}, and then apply a special saliency map, named Layer-CAM, to interpret the effectiveness of STA-MDCT in the time-frequency domain directly in Section \ref{subsec:interpret}.

\subsection{Saliency maps for speaker recognition}\label{subsec:saliency}
When making decisions, humans tend to focus on salient parts of an object and allocate their attention appropriately. Class activation map (CAM) is the saliency map of an image produced by a convolutional neural network (CNN) which emphasizes important regions for classifying the image. Several CAMs have been widely used in computer vision, such as the Grad-CAM \cite{selvaraju2017grad}, Grad-CAM++ \cite{chattopadhay2018grad}, Score-CAM \cite{wang2020score} and Layer-CAM \cite{jiang2021layercam}. In speech processing, Li \textit{et al.} \cite{li2022reliable} applied CAM to speaker recognition. Their study concludes that only Layer-CAM is a valid visualization tool for speaker recognition. This paper discusses Layer-CAM as well.

Here, we make a brief description to Layer-CAM in speaker recognition. Given a CNN-based speaker recognition system,
we denote the output feature maps of the final convolutional layer of the CNN as $\mathbf A$, and denote the $k$-th feature map in $\mathbf A$ as $\mathbf{A}^{k}$. Suppose the predicted score of the input $\mathbf{x}$ for the $c$-th speaker is:
\begin{equation}
    y^{c}=s(\mathbf{x}^{\mathrm{enroll}}_c, \mathbf{x}).
\end{equation}
We calculate the gradient of $y^{c}$ with respect to $\mathbf{A}^{k}$ by:
\begin{equation}
    w_{ij}^{k c}=\operatorname{relu}\left(\frac{\partial y^{c}}{\partial A_{ij}^{k}}\right).
\end{equation}
where $A_{ij}^{k}$ is the $(i,j)$-th location of $\mathbf{A}^{k}$
The saliency map $\mathbf{Z}^c$ of the input $\mathbf{x}$ at the location $(i,j)$ is:
 \begin{equation}
\label{equ:S}
    Z_{i j}^{c}=\operatorname{relu}\left\{\sum_{k} w_{i j}^{k c} \cdot A_{i j}^{k}\right\}.
\end{equation}
At last, we \textit{normalize} the saliency map $\mathbf{Z}^c$ by:
\begin{equation}
\label{equ:nor_S}
    \hat{\mathbf{Z}}^{c}=\frac{\mathbf{Z}^{c}-\min \mathbf{Z}^{c}}{\max \mathbf{Z}^{c}-\min \mathbf{Z}^{c}}.
\end{equation}

\begin{figure}[t]
\centering
\subfigure[ResNetSE34L]{
\begin{minipage}[t]{.95\linewidth}
\centering
\includegraphics[scale=0.25]{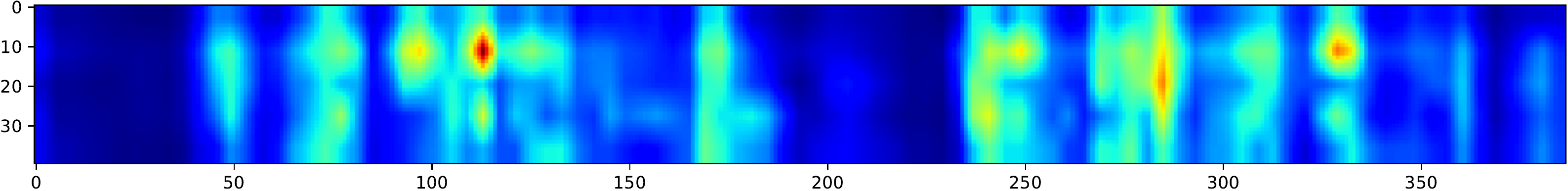}
\end{minipage}
}
\subfigure[ResNetSE34V2]{
\begin{minipage}[t]{.95\linewidth}
\centering
\includegraphics[scale=0.25]{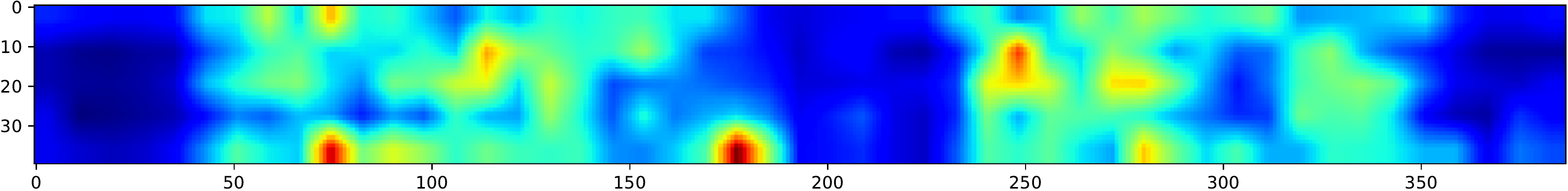}
\end{minipage}
}
\caption{Saliency maps for the speaker recognition models ResNetSE34L \cite{chung2020in} and ResNetSE34V2 \cite{kwon2021ins}. The regions with light and warm colors are critical regions indicating important time-frequency components in making decisions. }
\label{fig:vis}
\end{figure}

As illustrated in Fig. \ref{fig:vis}, given the same input $\mathbf{x}$, the saliency maps of different models for the same speaker $c$ significantly vary from each other, which clearly reveals that the models have different concerns on the same time-frequency unit.

\subsection{Interpretable STA-MDCT with Layer-CAM}
\label{subsec:interpret}

 Based on the observation in Section \ref{subsec:saliency}, if an attacker could successfully make a victim model shift its attention on the saliency map, then a successful attack may be made. Here we apply it to visually explain the effects of different transfer-based attackers on black-box victim models.

First, we select an utterance from the ground-truth speaker, and generate its adversarial voices by applying I-FGSM and STA-MDCT respectively to the ECAPA-TDNN \cite{desplanques2020ecapa} white-box surrogate model. Then, we apply the adversarial voices to attack the ResNetSE34L \cite{chung2020in} black-box victim model. Finally, we calculate the saliency maps of the above voices produced from ResNetSE34L, given the enrollment voices either from the ground-truth speaker or from the target speaker.

\begin{figure}[t]
\centering
\subfigure[Original test voice]{
\begin{minipage}[t]{.95\linewidth}
\centering
\includegraphics[scale=0.25]{raw_label_14.pdf}
\end{minipage}
\label{subfig:ori}
}
\subfigure[Adversarial test voice made by I-FGSM \cite{kurakin2016adversarial}]{
\begin{minipage}[t]{.95\linewidth}
\centering
\includegraphics[scale=0.25]{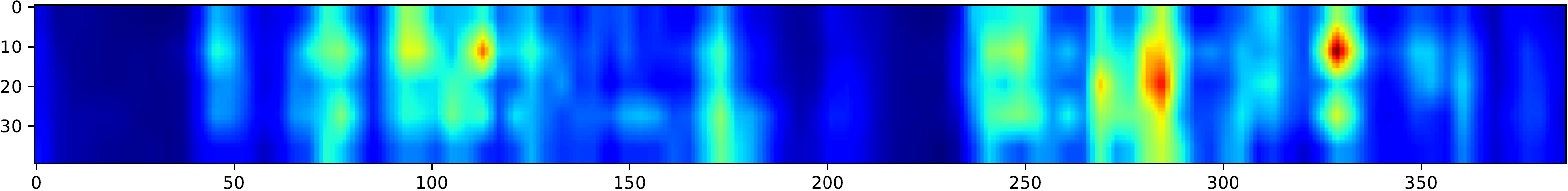}
\end{minipage}
\label{subfig:I-FGSM use_label}
}
\subfigure[Adversarial test voice made by STA-MDCT]{
\begin{minipage}[t]{.95\linewidth}
\centering
\includegraphics[scale=0.25]{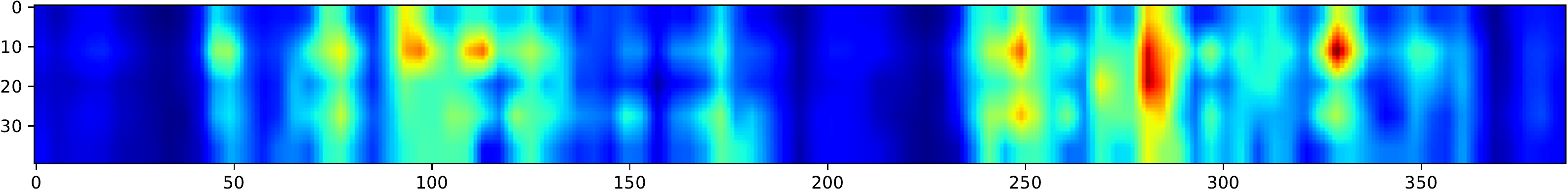}
\end{minipage}
\label{subfig:STA use_label}
}
\caption{Example of saliency maps, given the enrollment voice from the ground-truth speaker.}
\label{fig:use label}
\end{figure}

Fig. \ref{fig:use label} shows the saliency maps of the test voices, given the enrollment voice from the ground-truth speaker. Comparing Fig. \ref{fig:use label}a with Fig. \ref{fig:use label}b, we see that the attentive region of the saliency map of the adversarial voice generated by I-FGSM is quite similar with that generated from the original voice, which indicates that the transfer-based attack based on I-FGSM fails to shift the attention of the black-box victim model. On the contrary, comparing Fig. \ref{fig:use label}a with Fig. \ref{fig:use label}c, we see that STA-MDCT effectively shifts the victim model's attention from critical regions to other regions, which may lead to a classification error.

\begin{figure}[t]
\centering
\subfigure[Original test voice]{
\begin{minipage}[t]{.95\linewidth}
\centering
\includegraphics[scale=0.25]{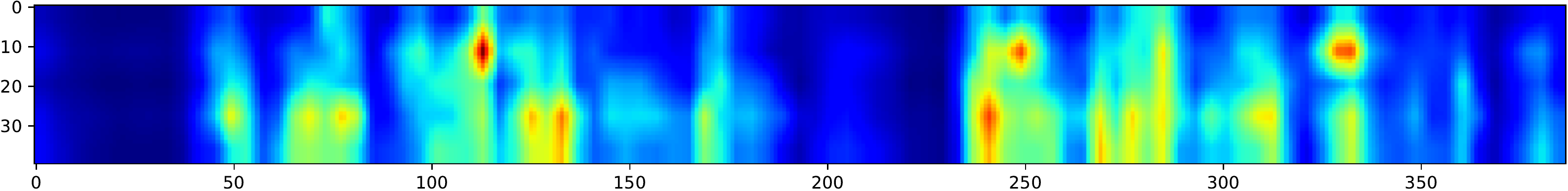}
\end{minipage}
}
\subfigure[Adversarial voice made by I-FGSM \cite{kurakin2016adversarial}]{
\begin{minipage}[t]{.95\linewidth}
\centering
\includegraphics[scale=0.25]{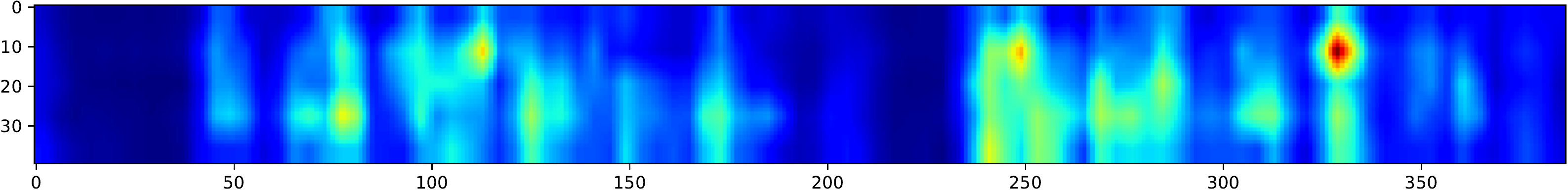}
\end{minipage}
}
\subfigure[Adversarial voice made by STA-MDCT]{
\begin{minipage}[t]{.95\linewidth}
\centering
\includegraphics[scale=0.25]{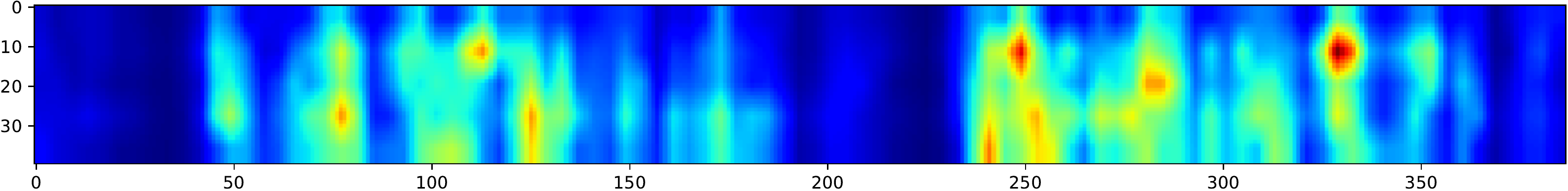}
\end{minipage}
}
\caption{Example of saliency maps, given the enrollment voice from the target speaker.}
\label{fig:use target}
\end{figure}

Fig. \ref{fig:use target} shows the saliency maps of the test voices, given the enrollment voice from the target speaker. Comparing Fig. \ref{fig:use target}a with Fig. \ref{fig:use label}a, we see that, when the victim model verifying the same test voice with different enrollment voices, the critical regions are shifted significantly. Comparing Fig. \ref{fig:use target}a with Fig. \ref{fig:use target}b, we see that the attentive region of the saliency map generated by I-FGSM is quite different from that of the original voice, which indicates that I-FGSM fails to make the targeted attack. On the other side, comparing Fig. \ref{fig:use target}a with Fig. \ref{fig:use target}c, we see that saliency maps of the original voice and the adversarial voice made by STA-MDCT share similar critical region, which may lead to a successful targeted attack.

\section{Experimental Setup}
\label{sec:Experimental Setup}
 In this section, we introduce the datasets, comparison attackers, evaluation metrics, and victim models.
\subsection{Datasets}

We first built four speaker recognition systems and then conducted adversarial attack experiments to them. See Section \ref{subsec:victim} for the detailed training process of the speaker recognition systems. The adversarial attack experiments were conducted on the VoxCeleb \cite{nagrani2017voxceleb} and LibriSpeech \cite{panayotov2015librispeech} datasets:

To attack an ASV system, we arbitrarily selected 1,000 trials from the \textit{Original-Clean} trial list of VoxCeleb1, which includes 500 target trials and 500 non-target trials. Then, an attacker transforms the clean test voices of all trials into adversarial examples which aim to make the ASV system yield opposite predictions from their ground-truth speaker identities.


For a speaker identification system, we first selected 10 speakers randomly, including 5 males and 5 females, from the \textit{test-other} and \textit{dev-other} subsets of LibriSpeech as the enrollment speakers, each of which contains 10 utterances.

To attack the speaker identification system in the CSI setting, we first picked 10 test speakers whose identities are the same as the enrollment speakers. Each of the speakers contains 100 randomly selected utterances that are \textit{different} from the enrollment data of the speaker. Then, for any of the 1,000 utterances, an attacker randomly picked a speaker identity from the 10 enrollment speakers that was different from the ground-truth identity of the utterance as the targeted label. Finally, it performed the targeted attack by generating an adversarial example from the clean test utterance, which aims to make the victim system wrongly predict the speaker identity as the targeted label.

To attack the speaker identification system in the OSI setting, we arbitrarily chose 10 test speakers that were different from the enrollment speakers from the \textit{train-other-500} subset of LibriSpeech. Each of the speakers contains 100 randomly selected utterances. Then, for any of the 1,000 utterances, an attacker randomly picked a speaker identity from the 10 enrollment speakers as the targeted label. Finally, it performed the targeted attack by generating an adversarial example from the clean test utterance, which aims to make the victim system wrongly predict the speaker identity as the targeted label.

 \subsection{Comparison adversarial attackers}
 The parameter setting of the proposed STA-MDCT was that $N = 20$, $\rho = 0.75$ and $\sigma = 44$ in all experiments unless otherwise stated.

We compare the proposed method with FGSM \cite{goodfellow2014explaining}, I-FGSM \cite{kurakin2016adversarial}, MI-FGSM \cite{dong2018boosting}, NI-FGSM \cite{lin2019nesterov} and ACG \cite{yamamura2022diversified}, see Section \ref{sec:baseline} for the descriptions of the comparison methods. All comparison methods were performed in their best settings or recommended default settings. That is,  the maximum perturbation $\epsilon = 40$, the iteration $T = 10$ and the step size $\alpha = \epsilon / T = 4$.

Both the white-box surrogate speaker recognition models and black-box victim models were selected from ResNetSE34L, ECAPA-TDNN, ResNetSE34V2 and RawNet3.



\subsection{Evaluation Metrics}

 The attacking effect of an attacker to a victim speaker recognition system  was evaluated in terms of the targeted attack success rate (TASR), false acceptance rate (FAR), equal error rate (EER), {identification error rate (IER)} and normalized minimum detection cost function (minDCF) with $P_{\mathrm{tar}}=0.05$ and $C_{\mathrm{miss}}=C_{\mathrm{fa}}=1$, produced from the victim system, where TASR refers to the proportion of the generated adversarial voices that are recognized as the targeted labels,  and {IER is the proportion of the input voices that are misclassified by the model.} The higher the evaluation scores are, the better the attacking effect is.

To measure the stealthiness of the adversarial examples, we used signal-to-noise (SNR), perceptual evaluation of speech quality (PESQ) \cite{941023}, and the standard $L_{2}$ norm. SNR is defined as $\operatorname{SNR}=10 \log _{10}\left(P_{\mathbf{x}}/P_{\mathbf{\delta}}\right)$ where $P_{\mathbf{x}}$ and $P_{\mathbf{\delta}}$ are the signal power of the benign voice $\mathbf{x}$ and the power of the perturbation $\mathbf{\delta}$ respectively. PESQ first applies an auditory transformation to obtain the loudness spectra of the benign voices and the adversarial voices, and then compares both loudness spectra to obtain a metric score with a value in the range of [-0.5, 4.5], see \cite{941023} for the details. Larger SNR, higher PESQ and smaller $L_{2}$ indicate better stealthiness.


\subsection{Victim speaker recognition systems}\label{subsec:victim}

First, we trained four representative speaker recognition systems, which are the ResNetSE34L \cite{chung2020in}, ECAPA-TDNN \cite{desplanques2020ecapa}, ResNetSE34V2 \cite{kwon2021ins} and RawNet3 \cite{jung2022pushing} respectively, on the development set of VoxCeleb2 \cite{chung2018voxceleb2}. Then, we applied the speaker recognition systems as the victim models for the tasks of the adversarial attack to ASV, CSI, and OSI, respectively. The parameter settings of the victim models are summarized as follows.

ResNetSE34L adopts attentive average pooling, and uses Angular Prototypical as the loss function\cite{chung2020in}. ECAPA-TDNN adopts attentive statistical pooling and AAM-Softmax \cite{liu2019large}. ResNetSE34V2 uses attentive statistical pooling, and takes the joint loss of the Angular Prototypical loss and softmax loss. RawNet3 uses attentive statistical pooling and AAM-Softmax.

In respect of the input acoustic features, RawNet3 uses raw waveforms as its input. The other three models first extract spectrograms with a hamming window of width 25ms and step size 10ms, and then apply log Mel-filterbanks to the spectrograms, followed by cepstral mean and variance normalization (CMVN).

To justify the advantage of the victim models, we evaluated them on the VoxCeleb1 \textit{Original-Clean} trial list which contains 37,000 trials from 40 speakers. The evaluation metrics include EER, minDCF, and IER. Table \ref{tab:spk_net} summarizes the speaker recognition performance of the models. From the table, we see that the models achieve the state-of-the-art performance.

Note that, the optimal decision thresholds $\theta$ were determined when the models were evaluated on the VoxCeleb1 \textit{Original-Clean} trial list. They were fixed thereafter, e.g. when we applied the attackers to the victim models.

\begin{table}[t]
\setlength\tabcolsep{4pt}
    \centering
    \caption{Performance of the speaker recognition models without attackers. The term ``{AP\_Softmax}'' refers to the joint loss function of {the Angular Prototypical loss and Softmax loss}.}
    \scalebox{0.78}{
    \begin{tabular}{cccccccc}
    \toprule
    \multirow{2}[2]{*}{Models}
    & \multicolumn{3}{c}{ASV}
    & \multicolumn{3}{c}{OSI}
    & CSI                   \\
    \cmidrule(r){2-4} \cmidrule(r){5-7} \cmidrule(r){8-8}
    \specialrule{0em}{2pt}{2pt}
    & EER (\%)   & MinDCF   &  $\theta$   & EER (\%) & IER (\%)  & $\theta$  & IER (\%)               \\
    \midrule
    ResNetSE34L \cite{chung2020in} +        & \multirow{2}{*}{2.179} & \multirow{2}{*}{0.168} & \multirow{2}{*}{0.051} & \multirow{2}{*}{2.8} & \multirow{2}{*}{0} & \multirow{2}{*}{0.72} & \multirow{2}{*}{0}  \\
    Angular Prototypical &                        &                        &                        &                      &                    &                       &                       \\
    \specialrule{0em}{2pt}{2pt}
    ECAPA-TDNN \cite{desplanques2020ecapa} +         & \multirow{2}{*}{1.172} & \multirow{2}{*}{0.08}  & \multirow{2}{*}{0.03}  & \multirow{2}{*}{1.0}   & \multirow{2}{*}{0} & \multirow{2}{*}{0.51} & \multirow{2}{*}{0}  \\
    AAM-Softmax          &                        &                        &                        &                      &                    &                       &                       \\
    \specialrule{0em}{2pt}{2pt}
    ResNetSE34V2 \cite{kwon2021ins} +       & \multirow{2}{*}{1.023} & \multirow{2}{*}{0.083} & \multirow{2}{*}{0.034} & \multirow{2}{*}{1.2}    & \multirow{2}{*}{0}  & \multirow{2}{*}{0.57}     & \multirow{2}{*}{0}     \\
    AP\_Softmax          &                        &                        &                        &                      &                    &                       &                       \\
    \specialrule{0em}{2pt}{2pt}
    RawNet3 \cite{jung2022pushing} +            & \multirow{2}{*}{1.039} & \multirow{2}{*}{0.069} & \multirow{2}{*}{0.03}  & \multirow{2}{*}{1.9}    & \multirow{2}{*}{0}  & \multirow{2}{*}{0.49}     & \multirow{2}{*}{0}     \\
    AAM-Softmax          &                        &                        &                        &                      &                    &                       &                       \\
    \bottomrule
    \end{tabular}
    }
    \label{tab:spk_net}
\end{table}

\section{Results of adversarial attacks to speaker verification}\label{subsec:ASV_result}
\label{sec:Result and Analysis}
In this section, we first report the comparison results between the STA-MDCT and the comparison attackers with a single white-box surrogate ASV model in Section \ref{subsec:single_STA_MDCT}, and with  an ensemble of white-box surrogate models in Section \ref{subsec:ensemble_STA_MDCT}. Then, we study the effects of the hyperparameters of STA-MDCT on performance in Section \ref{subsec:hyper}, and the effect of the SNR budget in Section \ref{subsec:SNR_budget} , given a single white-box surrogate ASV model.

\subsection{Results with a single white-box surrogate ASV model}\label{subsec:single_STA_MDCT}

\begin{table*}
  \centering
  \renewcommand\tabcolsep{3.5pt} 
  \caption{Performance of the comparison attackers that use the same white-box surrogate model, on the ASV task in terms of  EER (\%), TASR (\%), FAR (\%), and minDCF. The result in gray color indicates that it is a white-box attack where the victim model is also the surrogate model.
  }
  \scalebox{0.75}{
    \begin{tabular}{ccccccccccccccccccccc}
    \toprule
    \multirow{2}[2]{*}{Surrogate Model} & \multirow{2}[2]{*}{Attack} & \multicolumn{16}{c}{Victim Model}
    & \multicolumn{3}{c}{\multirow{2}[2]{*}{}} \\
          &       & \multicolumn{4}{c}{ResNetSE34L} & \multicolumn{4}{c}{ECAPA-TDNN} & \multicolumn{4}{c}{ResNetSE34V2} & \multicolumn{4}{c}{RawNet3}   &
          \multicolumn{3}{c}{}  \\
          \cmidrule(r){3-6} \cmidrule(r){7-10} \cmidrule(r){11-14} \cmidrule(r){15-18} \cmidrule(r){19-21}

          &       & \multicolumn{1}{c}{EER} & \multicolumn{1}{c}{TASR} & \multicolumn{1}{c}{FAR} & \multicolumn{1}{c}{MinDCF} & \multicolumn{1}{c}{EER} & \multicolumn{1}{c}{TASR} & \multicolumn{1}{c}{FAR} & \multicolumn{1}{c}{MinDCF} & \multicolumn{1}{c}{EER} & \multicolumn{1}{c}{TASR} & \multicolumn{1}{c}{FAR} & \multicolumn{1}{c}{MinDCF} & \multicolumn{1}{c}{EER} & \multicolumn{1}{c}{TASR} & \multicolumn{1}{c}{FAR} & \multicolumn{1}{c}{MinDCF} & \multicolumn{1}{c}{SNR(dB)} & \multicolumn{1}{c}{PESQ} & \multicolumn{1}{c}{L2} \\
    \midrule
    \multirow{7}[2]{*}{ResNetSE34L} & FGSM \cite{goodfellow2014explaining} & \textcolor{gray}{58.2} & \textcolor{gray}{59.5} & \textcolor{gray}{42.4} & \textcolor{gray}{1.00}   & 9.2 & 9.5 & 7.6 & 0.58 & 9.4 & 10.0  & 6.4 & 0.63 & 7.4 & 7.2 & 5.0   & 0.53 & 30.11 & 3.12 & 1.47 \\
        & I-FGSM \cite{kurakin2016adversarial} & \textcolor{gray}{97.2} & \textcolor{gray}{95.6} & \textcolor{gray}{92.6} & \textcolor{gray}{1.00}   & 16.0  & 20.5 & 8.8 & 0.73 & 18.6 & 24.2 & 7.2 & 0.81 & 9.8 & 13.0  & 6.0   & 0.70 & 37.02 & 4.13 & 0.63 \\
        & MI-FGSM \cite{dong2018boosting} & \textcolor{gray}{96.8} & \textcolor{gray}{95.8} & \textcolor{gray}{93.0}  & \textcolor{gray}{1.00}   & 22.6 & 26.3 & 10.8 & 0.89 & 25.6 & 30.1 & 11.6 & 0.96 & 16.2 & 19.4 & 8.0   & 0.84 & 32.61 & 3.59 & 1.09 \\
        & NI-FGSM \cite{lin2019nesterov} & \textcolor{gray}{\textbf{97.6}} & \textcolor{gray}{\textbf{96.5}} & \textcolor{gray}{\textbf{94.6}} & \textcolor{gray}{1.00}   & 23.8 & 27.1 & 11.4 & 0.90 & 26.6 & 31.6 & 12.0  & 0.96 & 17.0  & 20.6 & 8.8 & 0.85 & 32.65 & 3.55 & 1.10 \\
        & ACG \cite{yamamura2022diversified} & \textcolor{gray}{\textbf{97.6}} & \textcolor{gray}{94.2} & \textcolor{gray}{89.8} & \textcolor{gray}{1.00}   & 31.8 & 32.5 & 14.6 & 0.93 & 35.2 & 38.1 & 16.2 & 0.96 & 23.6 & 26.6 & 12.2 & 0.84 & 32.19 & 3.53 & 1.14 \\
        & STA-DCT {\cite{long2022frequency}} & \textcolor{gray}{90.6} & \textcolor{gray}{87.6} & \textcolor{gray}{80.2} & \textcolor{gray}{1.00}   & 32.0  & 33.2 & 16.4 & 0.95 & 32.0  & 34.9 & 16.0  & 0.96 & 26.6 & 28.2 & 12.8 & 0.94 & 34.65 & 3.92 & 0.84 \\
        & STA-MDCT & \textcolor{gray}{97.2} & \textcolor{gray}{96.1} & \textcolor{gray}{93.6} & \textcolor{gray}{1.00}   & \textbf{44.0} & \textbf{44.2} & \textbf{24.6} & 0.98 & \textbf{43.4} & \textbf{46.9} & \textbf{24.8} & 0.99 & \textbf{37.0} & \textbf{37.0} & \textbf{20.0}  & 0.98 & 34.18 & 3.93 & 0.89 \\
    \midrule
    \multirow{7}[2]{*}{ECAPA-TDNN} & FGSM \cite{goodfellow2014explaining} & 19.8 & 22.8 & 13.6 & 0.92 & \textcolor{gray}{62.0}  & \textcolor{gray}{59.5} & \textcolor{gray}{52.0}  & \textcolor{gray}{1.00}   & 13.2 & 15.7 & 8.6 & 0.83 & 18.0  & 20.0  & 13.0  & 0.91 & 30.12 & 3.10 & 1.47 \\
        & I-FGSM \cite{kurakin2016adversarial} & 32.0  & 35.6 & 20.6 & 0.98 & \textcolor{gray}{98.0}  & \textcolor{gray}{98.0}  & \textcolor{gray}{98.0}  & \textcolor{gray}{1.00}   & 26.4 & 29.8 & 13.4 & 0.96 & 40.0  & 38.9 & 21.8 & 1.00   & 36.30 & 4.12 & 0.68 \\
        & MI-FGSM \cite{dong2018boosting} & 40.2 & 45.7 & 25.4 & 0.99 & \textcolor{gray}{97.4} & \textcolor{gray}{97.6} & \textcolor{gray}{97.2} & \textcolor{gray}{1.00}   & 36.8 & 40.2 & 20.8 & 0.99 & 51.6 & 46.4 & 26.8 & 1.00   & 32.54 & 3.56 & 1.09 \\
        & NI-FGSM \cite{lin2019nesterov} & 38.6 & 44.5 & 25.2 & 0.99 & \textcolor{gray}{97.4} & \textcolor{gray}{97.7} & \textcolor{gray}{98.2} & \textcolor{gray}{1.00}   & 35.8 & 38.0  & 20.2 & 0.99 & 50.2 & 45.8 & 27.6 & 0.99 & 32.85 & 3.55 & 1.08 \\
        & ACG \cite{yamamura2022diversified} & 47.0  & 50.2 & 26.0  & 0.99 & \textcolor{gray}{\textbf{98.8}} & \textcolor{gray}{\textbf{98.7}} & \textcolor{gray}{\textbf{98.4}} & \textcolor{gray}{1.00}   & 49.0  & 49.5 & 27.4 & 0.99 & 64.8 & 59.2 & 42.0  & 1.00   & 31.88 & 3.49 & 1.18 \\
        & STA-DCT {\cite{long2022frequency}} & 49.4 & 53.6 & 31.0  & 1.00   & \textcolor{gray}{93.2} & \textcolor{gray}{92.6} & \textcolor{gray}{91.4} & \textcolor{gray}{1.00}   & 45.4 & 48.6 & 26.4 & 0.99 & 60.6 & 54.8 & 37.0  & 1.00   & 33.92 & 3.88 & 0.92 \\
        & STA-MDCT & \textbf{58.8} & \textbf{60.6} & \textbf{37.0}  & 1.00   & \textcolor{gray}{97.8}  & \textcolor{gray}{97.8}  & \textcolor{gray}{97.6}  & \textcolor{gray}{1.00}   & \textbf{58.2} & \textbf{56.5} & \textbf{33.8} & 1.00   & \textbf{72.2} & \textbf{68.5} & \textbf{54.0} & 1.00   & 33.43 & 3.84 & 0.97 \\
    \midrule
    \multirow{7}[2]{*}{ResNetSE34V2} & FGSM \cite{goodfellow2014explaining} & 14.4 & 16.7 & 11.2 & 0.82 & 9.2 & 9.6 & 7.8 & 0.58 & \textcolor{gray}{43.4} & \textcolor{gray}{44.0}  & \textcolor{gray}{29.2} & \textcolor{gray}{0.99} & 7.6 & 7.5 & 5.2 & 0.53 & 30.11 & 3.09 & 1.47 \\
        & I-FGSM \cite{kurakin2016adversarial} & 23.8 & 27.1 & 17.2 & 0.98 & 16.0  & 19.0  & 10.6 & 0.79 & \textcolor{gray}{96.8} & \textcolor{gray}{94.9} & \textcolor{gray}{92.0}  & \textcolor{gray}{1.00}   & 12.2 & 14.4 & 8.4 & 0.73 & 37.18 & 4.16 & 0.62 \\
        & MI-FGSM \cite{dong2018boosting} & 32.2 & 36.7 & 22.0  & 0.99 & 24.8 & 27.3 & 14.6 & 0.93 & \textcolor{gray}{96.6} & \textcolor{gray}{95.1} & \textcolor{gray}{92.6} & \textcolor{gray}{1.00}   & 19.4 & 22.7 & 11.4 & 0.86 & 32.71 & 3.60 & 1.08 \\
        & NI-FGSM \cite{lin2019nesterov} & 31.0  & 35.7 & 21.2 & 0.99 & 25.0  & 26.8 & 14.6 & 0.95 & \textcolor{gray}{96.2} & \textcolor{gray}{95.6} & \textcolor{gray}{93.2} & \textcolor{gray}{1.00}   & 19.4 & 22.4 & 11.6 & 0.88 & 33.10 & 3.59 & 1.06 \\
        & ACG \cite{yamamura2022diversified} & 34.4 & 38.5 & 20.0  & 0.99 & 30.4 & 32.0  & 17.2 & 0.93 & \textcolor{gray}{\textbf{97.8}} & \textcolor{gray}{\textbf{96.4}} & \textcolor{gray}{\textbf{93.8}} & \textcolor{gray}{1.00}   & 24.6 & 27.4 & 14.4 & 0.87 & 31.79 & 3.47 & 1.21 \\
        & STA-DCT {\cite{long2022frequency}} & 43.6 & 47.8 & 27.4 & 0.99 & 37.8 & 38.5 & 23.6 & 0.96 & \textcolor{gray}{89.0}  & \textcolor{gray}{87.9} & \textcolor{gray}{83.0}  & \textcolor{gray}{1.00}   & 31.8 & 35.3 & 20.8 & 0.98 & 34.69 & 3.97 & 0.84 \\
        & STA-MDCT & \textbf{54.8} & \textbf{56.9} & \textbf{35.2} & 1.00   & \textbf{51.2} & \textbf{49.9} & \textbf{31.8} & 1.00   & \textcolor{gray}{96.6} & \textcolor{gray}{94.9} & \textcolor{gray}{92.4} & \textcolor{gray}{1.00}   & \textbf{46.2} & \textbf{44.4} & \textbf{26.8} & 1.00   & 34.09 & 3.99 & 0.90 \\
    \midrule
    \multirow{7}[2]{*}{RawNet3} & FGSM \cite{goodfellow2014explaining} & 4.0   & 4.5 & 2.8 & 0.30 & 1.8 & 1.9 & 2.2 & 0.14 & 1.2 & 1.2 & 1.2 & 0.07 & \textcolor{gray}{2.6} & \textcolor{gray}{2.8} & \textcolor{gray}{2.0}   & \textcolor{gray}{0.21} & 30.11 & 3.14 & 1.47 \\
        & I-FGSM \cite{kurakin2016adversarial} & 4.0   & 4.1 & 4.4 & 0.30 & 2.6 & 2.4 & 2.8 & 0.17 & 1.6 & 1.5 & 1.6 & 0.11 & \textcolor{gray}{\textbf{19.4}} & \textcolor{gray}{19.6} & \textcolor{gray}{\textbf{14.8}} & \textcolor{gray}{0.90} & 40.00  & 4.19 & 0.47 \\
        & MI-FGSM \cite{dong2018boosting} & \textbf{5.2} & \textbf{5.2} & \textbf{4.8} & 0.34 & \textbf{3.4} & 3.4 & \textbf{3.2} & 0.20 & 1.6 & 1.9 & 2.2 & 0.15 & \textcolor{gray}{15.4} & \textcolor{gray}{17.0}  & \textcolor{gray}{12.2} & \textcolor{gray}{0.85} & 32.99 & 3.49 & 1.06 \\
        & NI-FGSM \cite{lin2019nesterov} & 4.4 & 5.1 & 4.2 & 0.31 & \textbf{3.4} & \textbf{3.5} & \textbf{3.2} & 0.20 & \textbf{2.0} & \textbf{2.0} & \textbf{2.4} & 0.15 & \textcolor{gray}{19.0}  & \textcolor{gray}{\textbf{20.5}} & \textcolor{gray}{14.6} & \textcolor{gray}{0.89} & 33.37 & 3.54 & 1.02 \\
        & ACG \cite{yamamura2022diversified} & 3.8 & 3.9 & 3.2 & 0.29 & 1.8 & 2.0   & 2.4 & 0.14 & 1.2 & 1.2 & 1.4 & 0.09 & \textcolor{gray}{4.4} & \textcolor{gray}{4.9} & \textcolor{gray}{3.8} & \textcolor{gray}{0.37} & 32.42 & 3.41 & 1.14 \\
        & STA-DCT {\cite{long2022frequency}} & 3.4 & 3.5 & 3.6 & 0.24 & 1.4 & 1.6 & 1.8 & 0.11 & 1.2 & 1.0   & 1.2 & 0.08 & \textcolor{gray}{3.2} & \textcolor{gray}{3.4} & \textcolor{gray}{3.0}   & \textcolor{gray}{0.26} & 40.09 & 4.20 & 0.46 \\
        & STA-MDCT & 3.0   & 2.9 & 3.0 & 0.19 & 1.4 & 1.4 & 1.8   & 0.09 & 1.0   & 1.0   & 1.2 & 0.09 & \textcolor{gray}{2.6} & \textcolor{gray}{2.5} & \textcolor{gray}{2.4} & \textcolor{gray}{0.18} & 40.09 & 4.24 & 0.47 \\
    \bottomrule
    \end{tabular}%
    }
  \label{tab:signal}%
\end{table*}%

Table \ref{tab:signal} lists the comparison result between the single-model attackers and the proposed STA-MDCT on the ASV task. From the comparison, we see that the proposed method consistently outperforms the comparison attackers. For example, the proposed STA-MDCT achieves relative EER improvement of 42.56\%, 44.18\%, and 31.1\%, respectively, over MI-FGSM, NI-FGSM and ACG. Particularly, adversarial examples generated with the ECAPA-TDNN surrogate ASV model tend to have better transferability to other black-box victim ASV models. Although adversarial examples generated with RawNet3 yield poor transferability to other victim models, the contrary transfer direction works fine, which indicates that adversarial examples generated from spectrum features are better than those generated from raw waves.

We should note that Table \ref{tab:signal} also lists the result of the white-box attacks in gray color, where the surrogate model and victim model are the same. From the result, we see that the attack performance of the proposed method is slightly weaker than ACG and NI-FGSM. It can be explained as that the proposed STA-MDCT, which improves the generalization ability of the adversarial examples to new black-box victim models, reduces the overfitting phenomenon of the adversarial examples to the surrogate models where they are generated.

\subsection{Results with an ensemble of white-box surrogate ASV models}\label{subsec:ensemble_STA_MDCT}

\begin{table*}
  \centering
  \renewcommand\tabcolsep{3.5pt} 
  \caption{Performance of the comparison attackers that use the same ensemble of white-box surrogate models, on the ASV task in terms of  EER (\%), TASR (\%), FAR (\%), and minDCF. The result in gray color indicates that it is a white-box attack where the victim model is one of the surrogate models.}
  \scalebox{0.75}{
    \begin{tabular}{ccccccccccccccccccccc}
    \toprule
    \multirow{2}[2]{*}{Surrogate Model} & \multirow{2}[2]{*}{Attack} & \multicolumn{16}{c}{Victim Model}
    & \multicolumn{3}{c}{\multirow{2}[2]{*}{}} \\
          &       & \multicolumn{4}{c}{ResNetSE34L} & \multicolumn{4}{c}{ECAPA-TDNN} & \multicolumn{4}{c}{ResNetSE34V2} & \multicolumn{4}{c}{RawNet3}   &
          \multicolumn{3}{c}{}  \\
          \cmidrule(r){3-6} \cmidrule(r){7-10} \cmidrule(r){11-14} \cmidrule(r){15-18} \cmidrule(r){19-21}

          &       & \multicolumn{1}{c}{EER} & \multicolumn{1}{c}{TASR} & \multicolumn{1}{c}{FAR} & \multicolumn{1}{c}{MinDCF} & \multicolumn{1}{c}{EER} & \multicolumn{1}{c}{TASR} & \multicolumn{1}{c}{FAR} & \multicolumn{1}{c}{MinDCF} & \multicolumn{1}{c}{EER} & \multicolumn{1}{c}{TASR} & \multicolumn{1}{c}{FAR} & \multicolumn{1}{c}{MinDCF} & \multicolumn{1}{c}{EER} & \multicolumn{1}{c}{TASR} & \multicolumn{1}{c}{FAR} & \multicolumn{1}{c}{MinDCF} & \multicolumn{1}{c}{SNR(dB)} & \multicolumn{1}{c}{PESQ} & \multicolumn{1}{c}{L2} \\
    \midrule
    \multicolumn{1}{c}{\multirow{7}[2]{*} {\makecell{ECAPA-TDNN \\ \& \\ ResNet34V2}}} & FGSM \cite{goodfellow2014explaining} & 21.8 & 25.0  & 15.6 & 0.98 & \textcolor{gray}{47.4} & \textcolor{gray}{46.6} & \textcolor{gray}{36.4} & \textcolor{gray}{0.99} & \textcolor{gray}{43.6} & \textcolor{gray}{43.9} & \textcolor{gray}{29.4} & \textcolor{gray}{1.00}   & 18.0  & 19.8 & 13.0  & 0.91 & 30.12 & 3.09 & 1.47 \\
        & I-FGSM \cite{kurakin2016adversarial} & 43.4 & 49.8 & 29.0  & 1.00   & \textcolor{gray}{95.8} & \textcolor{gray}{95.7} & \textcolor{gray}{95.2} & \textcolor{gray}{1.00}   & \textcolor{gray}{95.0}  & \textcolor{gray}{92.2} & \textcolor{gray}{87.4} & \textcolor{gray}{1.00}   & 48.8 & 45.4 & 25.4 & 1.00   & 36.37 & 4.12 & 0.68 \\
        & MI-FGSM \cite{dong2018boosting} & 48.2 & 53.3 & 31.4 & 1.00   & \textcolor{gray}{95.0}  & \textcolor{gray}{95.1} & \textcolor{gray}{94.6} & \textcolor{gray}{1.00}   & \textcolor{gray}{92.4} & \textcolor{gray}{90.2} & \textcolor{gray}{84.2} & \textcolor{gray}{1.00}   & 56.2 & 50.5 & 33.2 & 1.00   & 32.82 & 3.60 & 1.06 \\
        & NI-FGSM \cite{lin2019nesterov} & 46.0  & 52.8 & 32.6 & 1.00   & \textcolor{gray}{93.6} & \textcolor{gray}{93.0}  & \textcolor{gray}{91.0}  & \textcolor{gray}{1.00}   & \textcolor{gray}{95.0}  & \textcolor{gray}{93.0}  & \textcolor{gray}{89.4} & \textcolor{gray}{1.00}   & 51.2 & 47.2 & 29.0  & 1.00   & 33.19 & 3.60 & 1.05 \\
        & ACG \cite{yamamura2022diversified} & 55.2 & 56.6 & 32.4 & 1.00   & \textcolor{gray}{\textbf{97.2}} & \textcolor{gray}{\textbf{96.8}} & \textcolor{gray}{\textbf{96.2}} & \textcolor{gray}{1.00}   & \textcolor{gray}{\textbf{96.4}} & \textcolor{gray}{\textbf{94.1}} & \textcolor{gray}{\textbf{90.6}} & \textcolor{gray}{1.00}   & 67.0  & 61.3 & 42.8 & 1.00   & 31.85 & 3.49 & 1.19 \\
        & STA-DCT {\cite{long2022frequency}} & 56.6 & 58.5 & 36.0  & 1.00   & \textcolor{gray}{89.0}  & \textcolor{gray}{86.1} & \textcolor{gray}{81.6} & \textcolor{gray}{1.00}   & \textcolor{gray}{86.8} & \textcolor{gray}{83.6} & \textcolor{gray}{74.8} & \textcolor{gray}{1.00}   & 63.8 & 58.9 & 42.6 & 1.00   & 34.00  & 3.88 & 0.91 \\
        & STA-MDCT & \textbf{65.0} & \textbf{64.3} & \textbf{40.8} & 1.00   & \textcolor{gray}{96.0}  & \textcolor{gray}{95.8} & \textcolor{gray}{95.2} & \textcolor{gray}{1.00}   & \textcolor{gray}{95.4} & \textcolor{gray}{93.1} & \textcolor{gray}{88.8} & \textcolor{gray}{1.00}   & \textbf{73.6} & \textbf{70.5} & \textbf{56.0}  & 1.00   & 33.28 & 3.87 & 0.99 \\
    \midrule
    \multicolumn{1}{c}{\multirow{7}[2]{*} {\makecell{ResNetSE34L \\ \& \\ ResNet34V2}}} & FGSM \cite{goodfellow2014explaining} & \textcolor{gray}{49.6} & \textcolor{gray}{51.2} & \textcolor{gray}{36.6} & \textcolor{gray}{1.00}   & 12.6 & 13.8 & 9.6 & 0.68 & \textcolor{gray}{38.0}  & \textcolor{gray}{39.1} & \textcolor{gray}{25.4} & \textcolor{gray}{0.99} & 10.8 & 11.2 & 7.6 & 0.72 & 30.11 & 3.09 & 1.47 \\
        & I-FGSM \cite{kurakin2016adversarial} & \textcolor{gray}{94.2} & \textcolor{gray}{90.5} & \textcolor{gray}{83.4} & \textcolor{gray}{1.00}   & 29.6 & 32.4 & 15.6 & 0.94 & \textcolor{gray}{94.6} & \textcolor{gray}{92.1} & \textcolor{gray}{87.4} & \textcolor{gray}{1.00}   & 22.6 & 25.6 & 11.8 & 0.88 & 36.71 & 4.14 & 0.66 \\
        & MI-FGSM \cite{dong2018boosting} & \textcolor{gray}{94.0}  & \textcolor{gray}{90.9} & \textcolor{gray}{84.6} & \textcolor{gray}{1.00}   & 36.2 & 38.3 & 21.4 & 0.98 & \textcolor{gray}{92.2} & \textcolor{gray}{89.9} & \textcolor{gray}{84.0}  & \textcolor{gray}{1.00}   & 29.6 & 31.4 & 16.6 & 0.98 & 32.87 & 3.62 & 1.06 \\
        & NI-FGSM \cite{lin2019nesterov} & \textcolor{gray}{92.6} & \textcolor{gray}{87.5} & \textcolor{gray}{78.0}  & \textcolor{gray}{1.00}   & 36.4 & 39.3 & 21.6 & 0.97 & \textcolor{gray}{95.2} & \textcolor{gray}{\textbf{93.9}} & \textcolor{gray}{\textbf{91.2}} & \textcolor{gray}{1.00}   & 30.6 & 32.1 & 16.8 & 0.99 & 32.81 & 3.58 & 1.08 \\
        & ACG \cite{yamamura2022diversified} & \textcolor{gray}{95.4} & \textcolor{gray}{89.9} & \textcolor{gray}{81.8} & \textcolor{gray}{1.00}   & 46.6 & 46.2 & 26.2 & 0.99 & \textcolor{gray}{\textbf{96.6}} & \textcolor{gray}{93.5} & \textcolor{gray}{89.4} & \textcolor{gray}{1.00}   & 39.2 & 38.1 & 20.8 & 0.98 & 31.89 & 3.51 & 1.19 \\
        & STA-DCT {\cite{long2022frequency}} & \textcolor{gray}{86.2} & \textcolor{gray}{82.2} & \textcolor{gray}{71.4} & \textcolor{gray}{1.00}   & 48.4 & 48.4 & 29.6 & 1.00   & \textcolor{gray}{85.8} & \textcolor{gray}{82.1} & \textcolor{gray}{73.6} & \textcolor{gray}{1.00}   & 42.6 & 42.0  & 24.6 & 1.00   & 34.37 & 3.91 & 0.87 \\
        & STA-MDCT & \textcolor{gray}{\textbf{95.6}} & \textcolor{gray}{\textbf{93.4}} & \textcolor{gray}{\textbf{88.8}} & \textcolor{gray}{1.00}   & \textbf{61.6} & \textbf{57.6} & \textbf{38.8} & 1.00   & \textcolor{gray}{94.6} & \textcolor{gray}{92.2} & \textcolor{gray}{87.4} & \textcolor{gray}{1.00}   & \textbf{53.4} & \textbf{49.9} & \textbf{31.8} & 1.00   & 33.63 & 3.91 & 0.95 \\
    \midrule
    \multicolumn{1}{c}{\multirow{7}[2]{*} {\makecell{ResNetSE34L \\ \& \\ ECAPA-TDNN}}} & FGSM \cite{goodfellow2014explaining} & \textcolor{gray}{54.6} & \textcolor{gray}{56.0}  & \textcolor{gray}{39.8} & \textcolor{gray}{1.00}   & \textcolor{gray}{47.2} & \textcolor{gray}{46.3} & \textcolor{gray}{35.8} & \textcolor{gray}{1.00}   & 17.2 & 19.3 & 12.6 & 0.91 & 16.8 & 19.4 & 13.6 & 0.90 & 30.11 & 3.11 & 1.47 \\
        & I-FGSM \cite{kurakin2016adversarial} & \textcolor{gray}{94.6} & \textcolor{gray}{91.4} & \textcolor{gray}{84.6} & \textcolor{gray}{1.00}   & \textcolor{gray}{96.0}  & \textcolor{gray}{95.7} & \textcolor{gray}{95.4} & \textcolor{gray}{1.00}   & 39.8 & 43.4 & 21.6 & 0.99 & 46.4 & 42.5 & 22.8 & 1.00   & 36.25 & 4.11 & 0.69 \\
        & MI-FGSM \cite{dong2018boosting} & \textcolor{gray}{93.4} & \textcolor{gray}{89.2} & \textcolor{gray}{81.4} & \textcolor{gray}{1.00}   & \textcolor{gray}{94.8} & \textcolor{gray}{94.1} & \textcolor{gray}{93.4} & \textcolor{gray}{1.00}   & 45.0  & 47.6 & 27.6 & 1.00   & 51.4 & 47.0  & 30.2 & 1.00   & 32.80 & 3.60 & 1.07 \\
        & NI-FGSM \cite{lin2019nesterov} & \textcolor{gray}{93.6} & \textcolor{gray}{88.6} & \textcolor{gray}{79.4} & \textcolor{gray}{1.00}   & \textcolor{gray}{96.2} & \textcolor{gray}{96.3} & \textcolor{gray}{\textbf{96.2}} & \textcolor{gray}{1.00}   & 45.4 & 48.9 & 26.6 & 0.99 & 54.6 & 51.2 & 32.4 & 1.00   & 32.78 & 3.56 & 1.08 \\
        & ACG \cite{yamamura2022diversified} & \textcolor{gray}{95.6} & \textcolor{gray}{89.9} & \textcolor{gray}{81.8} & \textcolor{gray}{1.00}   & \textcolor{gray}{\textbf{97.4}} & \textcolor{gray}{\textbf{97.1}} & \textcolor{gray}{\textbf{96.2}} & \textcolor{gray}{1.00}   & 59.6 & 56.3 & 32.2 & 1.00   & 65.2 & 59.2 & 40.6 & 1.00   & 31.90 & 3.51 & 1.18 \\
        & STA-DCT {\cite{long2022frequency}} & \textcolor{gray}{88.0}  & \textcolor{gray}{83.2} & \textcolor{gray}{72.6} & \textcolor{gray}{1.00}   & \textcolor{gray}{87.6} & \textcolor{gray}{85.5} & \textcolor{gray}{80.4} & \textcolor{gray}{1.00}   & 50.4 & 52.4 & 30.0  & 1.00   & 59.4 & 53.7 & 36.6 & 1.00   & 34.00  & 3.87 & 0.91 \\
        & STA-MDCT & \textcolor{gray}{\textbf{95.8}} & \textcolor{gray}{\textbf{93.7}} & \textcolor{gray}{\textbf{88.9}} & \textcolor{gray}{1.00}   & \textcolor{gray}{95.6} & \textcolor{gray}{95.4} & \textcolor{gray}{94.6} & \textcolor{gray}{1.00}   & \textbf{65.6} & \textbf{60.8} & \textbf{38.8} & 1.00   & \textbf{69.8} & \textbf{66.8} & \textbf{51.6} & 1.00   & 33.34 & 3.84 & 0.98 \\
    \midrule
    \multicolumn{1}{c}{\multirow{7}[2]{*} {\makecell{ResNetSE34L \\ \& \\ ECAPA-TDNN \\ \& \\ ResNet34V2}}} & FGSM \cite{goodfellow2014explaining} & \textcolor{gray}{49.4} & \textcolor{gray}{51.9} & \textcolor{gray}{36.2} & \textcolor{gray}{1.00}   & \textcolor{gray}{43.3} & \textcolor{gray}{41.2} & \textcolor{gray}{29.4} & \textcolor{gray}{0.98} & \textcolor{gray}{39.8} & \textcolor{gray}{41.4} & \textcolor{gray}{27.2} & \textcolor{gray}{1.00}   & 18.8 & 20.5 & 14.2 & 0.91 & 30.11 & 3.09 & 1.47 \\
        & I-FGSM \cite{kurakin2016adversarial} & \textcolor{gray}{93.0}  & \textcolor{gray}{88.3} & \textcolor{gray}{79.2} & \textcolor{gray}{1.00}   & \textcolor{gray}{94.6} & \textcolor{gray}{94.1} & \textcolor{gray}{93.0}  & \textcolor{gray}{1.00}   & \textcolor{gray}{93.4} & \textcolor{gray}{90.6} & \textcolor{gray}{84.6} & \textcolor{gray}{1.00}  & 53.4 & 48.3 & 29.0  & 1.00   & 36.21 & 4.11 & 0.69 \\
        & MI-FGSM \cite{dong2018boosting} & \textcolor{gray}{92.2} & \textcolor{gray}{87.5} & \textcolor{gray}{78.2} & \textcolor{gray}{1.00}   & \textcolor{gray}{93.8} & \textcolor{gray}{92.9} & \textcolor{gray}{91.2} & \textcolor{gray}{1.00}   & \textcolor{gray}{91.2} & \textcolor{gray}{87.6} & \textcolor{gray}{80.2} & \textcolor{gray}{1.00}   & 57.2 & 52.6 & 35.8 & 1.00   & 32.33 & 3.56 & 1.12 \\
        & NI-FGSM \cite{lin2019nesterov} & \textcolor{gray}{92.8} & \textcolor{gray}{90.2} & \textcolor{gray}{\textbf{84.2}} & \textcolor{gray}{1.00}   & \textcolor{gray}{90.2} & \textcolor{gray}{87.7} & \textcolor{gray}{82.8} & \textcolor{gray}{1.00}   & \textcolor{gray}{91.2} & \textcolor{gray}{88.3} & \textcolor{gray}{81.4} & \textcolor{gray}{1.00}   & 50.6 & 46.7 & 28.2 & 1.00   & 33.37 & 3.62 & 1.04 \\
        & ACG \cite{yamamura2022diversified} & \textcolor{gray}{\textbf{94.6}} & \textcolor{gray}{87.9} & \textcolor{gray}{78.2} & \textcolor{gray}{1.00}   & \textcolor{gray}{\textbf{96.4}} & \textcolor{gray}{\textbf{95.5}} & \textcolor{gray}{\textbf{94.2}} & \textcolor{gray}{1.00}   & \textcolor{gray}{\textbf{96.0}} & \textcolor{gray}{\textbf{92.5}} & \textcolor{gray}{\textbf{87.6}} & \textcolor{gray}{1.00}   & 68.2 & 62.0  & 42.6 & 1.00   & 31.86 & 3.51 & 1.19 \\
        & STA-DCT {\cite{long2022frequency}} & \textcolor{gray}{86.8} & \textcolor{gray}{82.2} & \textcolor{gray}{70.8} & \textcolor{gray}{1.00}   & \textcolor{gray}{89.8} & \textcolor{gray}{88.6} & \textcolor{gray}{84.6} & \textcolor{gray}{1.00}   & \textcolor{gray}{87.8} & \textcolor{gray}{83.3} & \textcolor{gray}{74.4} & \textcolor{gray}{1.00}   & 71.6 & 67.2 & 51.8 & 1.00   & 32.59 & 3.71 & 1.09 \\
        & STA-MDCT & \textcolor{gray}{94.4} & \textcolor{gray}{\textbf{90.6}} & \textcolor{gray}{83.2} & \textcolor{gray}{1.00}   & \textcolor{gray}{94.8} & \textcolor{gray}{93.9} & \textcolor{gray}{92.4} & \textcolor{gray}{1.00}   & \textcolor{gray}{93.4} & \textcolor{gray}{90.8} & \textcolor{gray}{85.0}  & \textcolor{gray}{1.00}   & \textbf{72.8} & \textbf{70.1} & \textbf{54.8} & 1.00   & 33.27 & 3.85 & 0.99 \\
    \bottomrule
    \end{tabular}%
    }
  \label{tab:ensemble}%
\end{table*}%

Crafting adversarial examples from an ensemble of surrogate models has been shown to be an effective way in improving the transferability of an attacker. Here we conducted an experimental comparison of the attackers with an ensemble of surrogate models that were selected from the four ASV models. From the results in Table \ref{tab:ensemble}, we observe that our method achieves the highest EER and TASR in all black-box attack scenarios. For instance, the proposed method with the ensemble of the ResNetSE34L and ResNetSE34V2 surrogate models achieves an EER of 61.6\% on the ECAPA-TDNN victim model, which is absolutely 10\% higher than the result of the corresponding single white-box surrogate model in Table \ref{tab:signal}.

\subsection{Ablation study}\label{subsec:hyper}
\label{Sec:hyper-p}

\begin{figure*}[t]
\begin{center}
\includegraphics[width=0.98\textwidth]{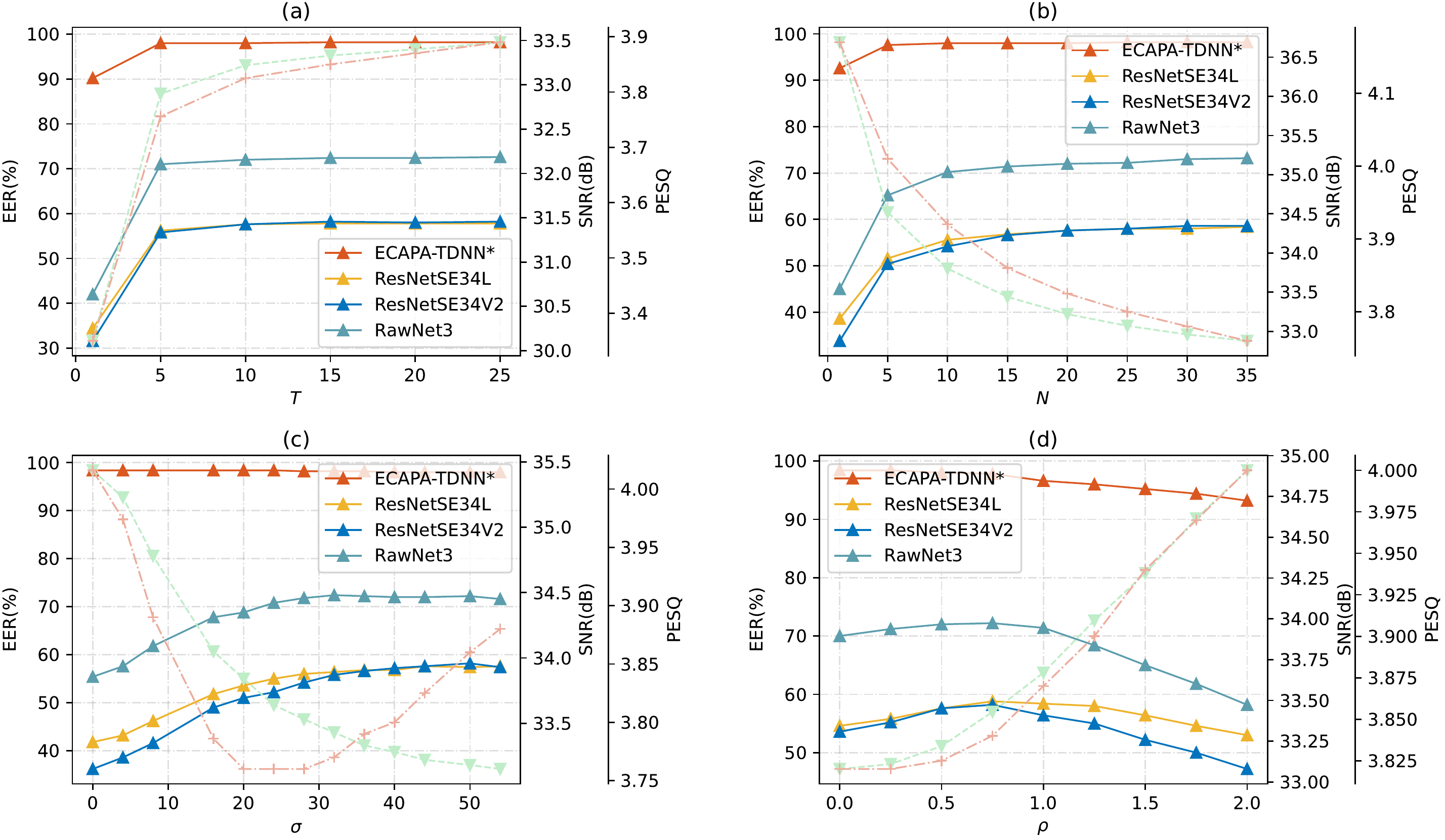}
\end{center}
\caption{Effects of the hyperparameters of STA-MDCT on performance in terms of EER (solid line), SNR (dashed line), and PESQ (dotted line). ECAPA-TDNN is used as the white-box surrogate model. The marker ``*'' indicates the white-box attack.}
\label{fig:tune_param}
\end{figure*}


In this subsection, we investigate the effects of the hyperparameters of STA-MDCT, including the maximum iterations $T$, number of spectrum transformation $N$, standard deviation $\boldsymbol{\sigma}$ of noise $\boldsymbol{\xi}$, and tuning factor $\boldsymbol{\rho}$. For tuning each hyperparameter, we fixed the others to their default values. We crafted adversarial examples from the ECAPA-TDNN white-box surrogate model, and apply them to attack the remaining three black-box victim models. The results are summarized in Fig. \ref{fig:tune_param} and analyzed as follows. Note that, we also show the result of the white-box attack as a reference.

\subsubsection{Effect of the maximum iterations $T$}
From Fig. \ref{fig:tune_param}a, we see that, when $T = 1$, the EER and PESQ produced by the proposed method is far from satisfactory; as $T$ increases, both the transferability and the SNR/PESQ of the adversarial examples improves, with a negative effect of the increased computational cost. To balance the two factors, we set $T=10$ as the default.
\subsubsection{Effect of the spectrum transformation $N$}
From Fig. \ref{fig:tune_param}b, we see that, when $N = 1$, a single spectrum transformation yields the worst transferability; the transferability of the adversarial examples is improved when $N$ increases, and tends to be increased slowly when $N$ exceeds 20. This phenomenon indicates that the proposed spectrum transformation can effectively narrow the gap between the white-box surrogate model and the black-box victim models. To balance the computational cost, we set $N=20$ in this paper.
\subsubsection{Effect of the standard deviation $\sigma$}
From Fig. \ref{fig:tune_param}c, we see that the larger $\sigma$ is, the higher the magnitude of the adversarial noise will be. Therefore, the EERs of the black-box victim models first increase dramatically, and then drop slightly; the highest EERs appear around $\sigma=44$. On the other side, as $\sigma$ increases, the SNR of the adversarial examples decreases continuously; the PESQ first decreases substantially, and then gradually increases. To balance the above three factors, we choose $\sigma=44$ in this paper.

\subsubsection{Effect of the tuning factor $\rho$}
From Fig. \ref{fig:tune_param}d, we see that, as $\rho$ increases, the EER curves of the black-box victim models gradually reach the peak at around $\rho= 0.75$. As $\rho$ continues to increase, the EER curves decrease due to the excessive spectral transformation. On the other side, the SNR and PESQ are increased constantly with the increase of $\rho$. Consequently, we choose $\rho=0.75$ in this paper.

\subsection{Effect of the SNR budget on performance}\label{subsec:SNR_budget}

All of the above experiments were conducted by setting the perturbation level $\epsilon=40$, where the SNRs of the adversarial examples were around 33 dB. In this section, we study the attack effect under the situation that the SNR is controlled to be larger than a given threshold $b$, named the \textit{SNR budget}. To generate a large number of adversarial examples with various SNR levels, we set $\epsilon$ to a wide range of $\{5, 10, 20, 30, 40, 50\}$. The white-box surrogate model was ECAPA-TDNN. The black-box victim model was ResNetSE34L.

\begin{figure*}[t]
\begin{center}
\includegraphics[width=0.8\textwidth]{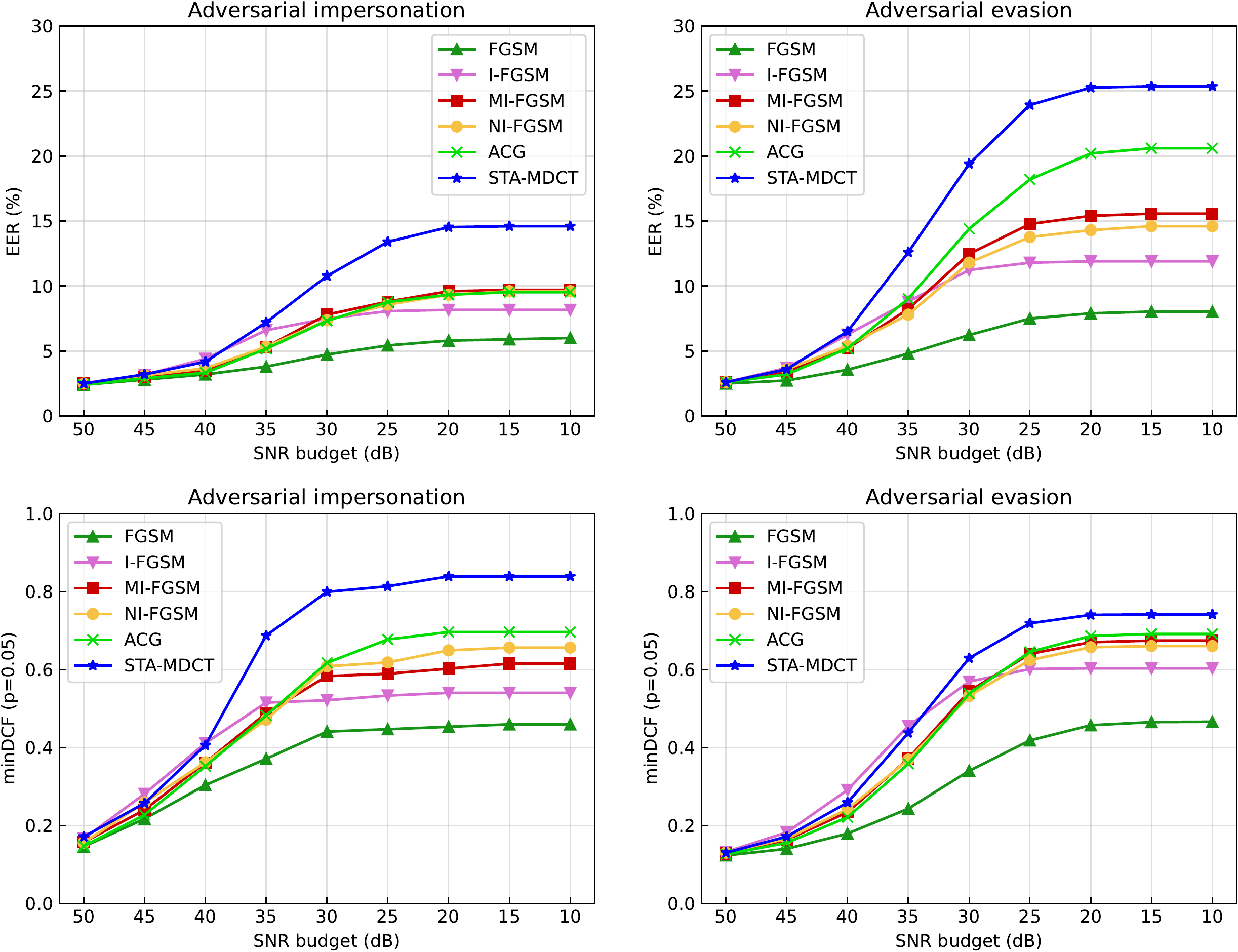}
\end{center}
\caption{Performance of the comparison methods with respect to the SNR budget. The black-box victim model is ResNetSE34L. The white-box surrogate model is ECAPA-TDNN.}
\label{fig:snr_budget}
\end{figure*}

\begin{figure}[t]
\begin{center}
\includegraphics[width=0.35\textwidth]{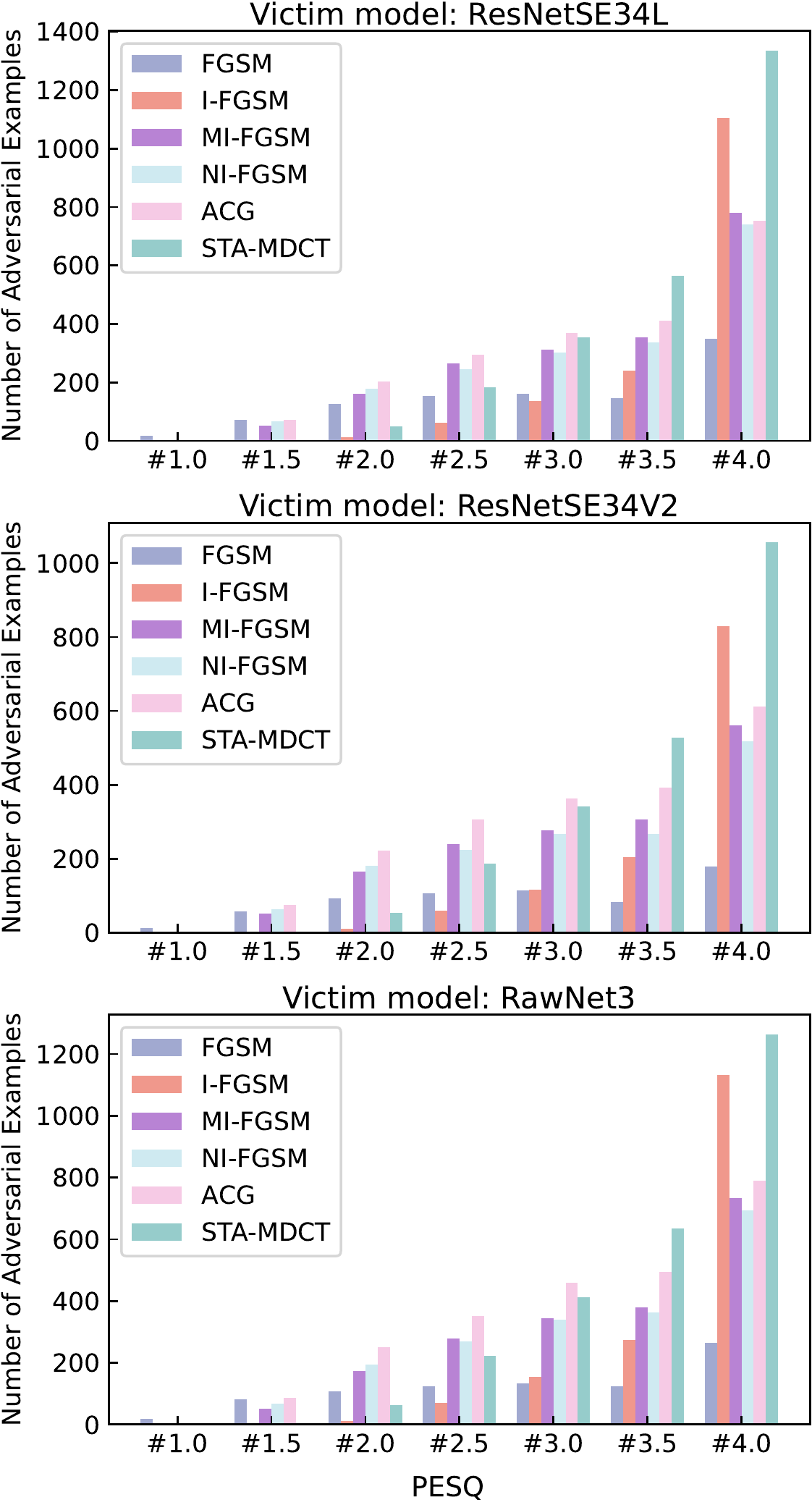}
\end{center}
\caption{Histograms of the adversarial examples that can successfully deceive the target victim model in terms of PESQ. The symbol ``\#i" refers to the range ``[i, i+0.5)" in PESQ.}
\label{fig:PESQ_budget}
\end{figure}

To study the effect of the SNR budget $b$ on performance, we count the EER and minDCF statistics of the impersonation attacks and evasion attacks separately as in \cite{villalba2020x, dong2020benchmarking}. Specifically, given an original trial set $\mathcal{O}=\left\{\left(\mathbf{x}_{i}^{\mathrm{enroll}}, \mathbf{x}_{i}^{\mathrm{test}}\right) \mid i=1,2, \cdots, I\right\}$ and its corresponding adversarial trial set  $\mathcal{A}=\left\{\left(\mathbf{x}_{i}^{\mathrm{enroll}}, \mathbf{x}_{i}^{\mathrm{adv}}\right) \mid i=1,2, \cdots, I\right\}$.
Suppose $\mathbf{p}_{\mathrm{adv}}=\left[p_{\mathrm{adv}, 1}, \ldots, p_{\mathrm{adv}, I}\right]^{T}$ is a vector describing the SNRs of the adversarial trials. For a given SNR budget $b$, we can obtain a mixed trial set $\mathcal{M}(b)$ whose elements are defined by:
\begin{align}
    {t}_{i}=\left\{\begin{array}{ll}
    (\mathbf{x}_{i}^{\mathrm{enroll}}, \mathbf{x}_{i}^{\mathrm{adv}}), & \mbox{ if } p_{\mathrm{adv}, i} \geq b \mbox{ and }i \in {G} \\
    (\mathbf{x}_{i}^{\mathrm{enroll}}, \mathbf{x}_{i}^{\mathrm{test}}), & \mbox { otherwise }
\end{array} \right.\nonumber\\
\forall i = 1,\ldots,I.\nonumber
\end{align}
where $G$ is defined as the set of non-target trials when the task is the impersonation attack to ASV, and defined as the set of target trials when the task is the evasion attack to ASV. Finally, the EER and minDCF are calculated from the mixed trial set $\mathcal{M}(b)$.

 Fig. \ref{fig:snr_budget} shows the EER and minDCF of the victim model with respect to the SNR budget $b$, where we have separately summarized the impersonation and evasion attacks. From the figure, we can observe that (i) as the SNR budget $b$ decreases, the EER of the victim model increases for all comparison attackers; (ii) when the SNR budget is below 40dB, the proposed method achieves a significantly higher EER than the other comparison methods; (iii) when the SNR budget is above 40dB, the proposed method achieves an EER comparable to I-FGSM, and outperforms the other comparison methods. The experimental phenomena in minDCF are similar to those in EER.

Additionally, we also analyzed the PESQ of the successful attacks of the above adversarial examples. Fig. \ref{fig:PESQ_budget} shows the number of the \textit{successful} adversarial attacks in different PESQ ranges, where the adversarial attacks that were not successful in deceiving the target victim model were discarded. From the figure, we see that the the PESQ of the successful adversarial examples generated by the proposed method is higher than all comparison methods, which provides a strong evidence that the adversarial perturbations generated by the proposed method are more imperceptible to human.

\section{Results of adversarial attacks to speaker identification}\label{subsec:si_result}

\begin{table}[t]
  \centering
  \caption{Performance of the comparison attackers on the CSI task in terms of IER (\%) and TASR (\%), where the average SNR of adversarial examples is 33.5dB. The marker ``*'' indicates the white-box attack.}
  \scalebox{0.75}{
    \begin{tabular}{cccccccccc}
    \toprule
    \multirow{3}[4]{*}{Attack} & \multicolumn{8}{c}{Victim Model}              &  \\
        & \multicolumn{2}{c}{ECAPATDNN*} & \multicolumn{2}{c}{ResNetSE4L} & \multicolumn{2}{c}{ResNetSE34V2} & \multicolumn{2}{c}{RawNet3} & \multirow{2}[3]{*}{PESQ} \\
\cmidrule{2-9}        & IER & TASR & IER & TASR & IER & TASR & IER & TASR &  \\
    \midrule
    FGSM \cite{goodfellow2014explaining} & 5.7 & 4.7 & 1.9 & 0.6 & 0.1 & 0.1 & 90.0  & 9.4 & 2.74 \\
    I-FGSM \cite{kurakin2016adversarial} & \textbf{100.0} & \textbf{100.0} & 9.9 & 8.2 & 1.4 & 1.4 & 90.0  & 9.2 & 3.07 \\
    MI-FGSM \cite{dong2018boosting} & 99.8 & 99.8 & 9.6 & 7.7 & 2.0   & 2.0   & 90.0  & \textbf{11.0}  & 2.94 \\
    NI-FGSM \cite{lin2019nesterov} & \textbf{100.0} & \textbf{100.0} & 9.4 & 7.6 & 1.3 & 1.3 & 90.0  & 9.6 & 2.91 \\
    ACG \cite{yamamura2022diversified} & 99.9 & 99.9 & 7.0   & 5.8 & 0.8 & 0.7 & \textbf{92.2} & 10.3 & 2.98 \\
    STA-MDCT & 99.0  & 98.9 & \textbf{39.1} & \textbf{27.7} & \textbf{18.8} & \textbf{17.8} & 90.0  & 10.1 & 3.16 \\
    \bottomrule
    \end{tabular}%
    }
  \label{tab:csi}%
\end{table}%

\begin{table}[t]
  \centering
  \caption{Performance of the comparison attackers on the OSI task in terms of IER (\%) and TASR (\%), where the average SNR of adversarial examples is 33.5dB. The marker ``*'' indicates the white-box attack.}
  \scalebox{0.75}{
    \begin{tabular}{cccccccccc}
    \toprule
    \multirow{3}[4]{*}{Attack} & \multicolumn{8}{c}{Victim Model}              &  \\
        & \multicolumn{2}{c}{ECAPATDNN*} & \multicolumn{2}{c}{ResNetSE4L} & \multicolumn{2}{c}{ResNetSE34V2} & \multicolumn{2}{c}{RawNet3} & \multirow{2}[3]{*}{PESQ} \\
\cmidrule{2-9}        & IER & TASR & IER & TASR & IER & TASR & IER & TASR &  \\
    \midrule
    FGSM \cite{goodfellow2014explaining} & 19.5 & 18.1 & 4.9 & 2.1 & 1.5 & 1.0   & 0.0   & 0.0   & 3.06 \\
    I-FGSM \cite{kurakin2016adversarial} & \textbf{100.0} & \textbf{100.0} & 14.4 & 9.2 & 7.4 & 6.0   & 0.0   & 0.0   & 3.29 \\
    MI-FGSM \cite{dong2018boosting} & 99.4 & 99.4 & 11.6 & 7.3 & 6.5 & 4.6 & 0.0   & 0.0   & 3.32 \\
    NI-FGSM \cite{lin2019nesterov} & \textbf{100.0} & \textbf{100.0} & 13.1 & 7.9 & 6.4 & 5.0   & 0.0   & 0.0   & 3.2 \\
    ACG \cite{yamamura2022diversified} & 99.9 & 99.9 & 12.5 & 6.9 & 6.3 & 4.5 & 0.0   & 0.0   & 3.26 \\
    STA-MDCT & 99.5 & 99.5 & \textbf{23.2} & \textbf{19.0}  & \textbf{16.7} & \textbf{15.8} & 0.0   & 0.0   & 3.35 \\
    \bottomrule
    \end{tabular}%
    }
  \label{tab:osi}%
\end{table}%

Tables \ref{tab:csi} and \ref{tab:osi} list the performance of the comparison attackers in the CSI and OSI scenarios, respectively, where we combine each attacker with the single white-box surrogate model ECAPA-TDNN, and then apply the generated adversarial examples to the other three black-box victim systems. From the tables, we see that, when the SNR budget is controlled to be the same around 33.5dB, the proposed STA-MDCT achieves better attack performance in terms of IER and TASR, as well as higher PESQ than the comparison methods. In addition, the black-box attack performance of all comparison methods is generally poor. It may be caused by the following reasons. First, the dataset used to attack the black-box victim models is different from that of the white-box surrogate model. Second, attacking speaker identification refers to many enrolled speaker identities, making the attack more challenging than attacking ASV. Besides, compared to attacking a CSI system, attacking an OSI system not only needs to maximize the confidence score for determining an adversarial voice to a target label, but also has to make the score exceed the predefined threshold $\theta$.

\section{Conclusions}
\label{sec:Conclusions}
In this paper, we propose a spectrum transformation attack method based on modified discrete cosine transform. {It first applies MDCT to the input voices and then slightly modifies the energy of the frequency bands of the transformed voices in the time-frequency domain for capturing the salient regions of the adversarial noise that are critical to a successful attack.}
 Different from existing transfer-based attackers, STA-MDCT generates adversarial examples in the time-frequency domain, which improves the transferability and efficiency of the adversarial attack. Moreover, we also interpret the effectiveness of transfer-based attacks by Layer-CAM. To our knowledge, it is the first study that the transferability of transfer-based attackers is observable directly from the critical attention regions of saliency maps. The comprehensive experiments on the ASV, OSI and CSI tasks demonstrate the effectiveness of the proposed method.

\bibliographystyle{IEEEtran}
\end{document}